\documentclass[twocolumn, trackchanges]{aastex701}

\usepackage{booktabs}
\usepackage{amsmath}

\begin{document}

\title{A Pair of Warm Saturn-mass Planets near the 2:1 Mean Motion Resonance around TOI-3850}
\shorttitle{Two Warm Saturns around TOI-3850}
\shortauthors{Collins et al.}

\correspondingauthor{Sean Collins}

\author[0009-0002-9657-6874]{Sean Collins}
\email[show]{scolli05@student.ubc.ca} 
\affiliation{Department of Physics and Astronomy, University of British Columbia, 6224 Agricultural Road, Vancouver, BC V6T 1Z1, Canada}

\author[0000-0001-9269-8060]{Michelle Kunimoto}
\affiliation{Department of Physics and Astronomy, University of British Columbia, 6224 Agricultural Road, Vancouver, BC V6T 1Z1, Canada}
\email{mkuni@phas.ubc.ca}

\author[0000-0001-6129-5699]{Nicolas B. Cowan}
\email{nicolas.cowan@mcgill.ca} 
\affiliation{Department of Physics, McGill University, 3600 rue University, Montréal, QC H3A 2T8, Canada}
\affiliation{Department of Earth \& Planetary Sciences, McGill University, 3450 rue University, Montreal, QC H3A 0E8, Canada}

\author[0000-0002-3481-9052]{Keivan G. Stassun}
\email{keivan.stassun@Vanderbilt.edu} 
\affiliation{Department of Physics and Astronomy, Vanderbilt University, Nashville, TN 37235, USA}

\author[0000-0001-6513-1659]{Jack J. Lissauer}
\email{jack.lissauer@nasa.gov} 
\affiliation{NASA Ames Research Center, Moffett Field, CA 94035, USA}

\author{Ze’ev Vladimir}
\email{zvladimir@uchicago.edu} 
\affiliation{Department of Astronomy \& Astrophysics, University of Chicago, 5640 S. Ellis Avenue, Chicago, IL 60637, USA}

\author[0000-0003-4603-556X]{Teo Mo\v{c}nik}
\email{teo.mocnik@noirlab.edu}
\affiliation{Gemini Observatory/NSF NOIRLab, 670 N. A'ohoku Place, Hilo, HI 96720, USA}

\author[0009-0005-1609-8713]{Ernesto Elenter}
\email{eelenter@gmail.com} 
\affiliation{Instituto de Física, Facultad de Ciencias, UDELAR, Igua 4225, Montevideo 11400, Uruguay}

\author[0000-0001-9911-7388]{David W. Latham}
\email{dlatham@cfa.harvard.edu} 
\affiliation{Center for Astrophysics \textbar \ Harvard \& Smithsonian, 60 Garden Street, Cambridge, MA 02138, USA}

\author{Karen A. Collins}
\email{karen.collins@cfa.harvard.edu} 
\affiliation{Center for Astrophysics \textbar \ Harvard \& Smithsonian, 60 Garden Street, Cambridge, MA 02138, USA}

\author{Jacob Bean}
\email{jacobbean@uchicago.edu} 
\affiliation{Department of Astronomy \& Astrophysics, University of Chicago, 5640 S. Ellis Avenue, Chicago, IL 60637, USA}

\author[0009-0008-5145-0446]{Stephanie Striegel}
\email{stephanie.l.striegel@nasa.gov} 
\affiliation{NASA Ames Research Center, Moffett Field, CA 94035, USA}

\author[0000-0003-1464-9276]{Khalid Barkaoui}
\email{khalid.barkaoui@uliege.be } 
\affiliation{Instituto de Astrof\'isica de Canarias (IAC), Calle V\'ia L\'actea s/n, 38200, La Laguna, Tenerife, Spain}
\affiliation{Astrobiology Research Unit, Universit\'e de Li\`ege, 19C All\'ee du 6 Ao\^ut, 4000 Li\`ege, Belgium}
\affiliation{Department of Earth, Atmospheric and Planetary Science, Massachusetts Institute of Technology, 77 Massachusetts Avenue, Cambridge, MA 02139, USA}

\author[0000-0003-4508-2436]{Ritvik Basant}
\email{rbasant@uchicago.edu} 
\affiliation{Department of Astronomy \& Astrophysics, University of Chicago, 5640 S. Ellis Avenue, Chicago, IL 60637, USA}

\author[0009-0005-1486-8374]{Tanya Das}
\email{tanyadas@uchicago.edu}
\affiliation{Department of Astronomy \& Astrophysics, University of Chicago, 5640 S. Ellis Avenue, Chicago, IL 60637, USA}

\author[0000-0002-6482-2180]{Raquel For\'es-Toribio}
\email{forestoribio.1@osu.edu} 
\affiliation{Department of Astronomy, The Ohio State University, 140 West 18th Avenue, Columbus, OH 43210, USA}
\affiliation{Center for Cosmology and Astroparticle Physics, The Ohio State University, 191 W. Woodruff Avenue, Columbus, OH 43210, USA}

\author[0000-0002-4909-5763]{Akihiko Fukui}
\email{afukui@g.ecc.u-tokyo.ac.jp} 
\affiliation{Komaba Institute for Science, The University of Tokyo, 3-8-1 Komaba, Meguro, Tokyo 153-8902, Japan}
\affiliation{Instituto de Astrof\'isica de Canarias (IAC), Calle V\'ia L\'actea s/n, 38200, La Laguna, Tenerife, Spain}

\author[0000-0001-9833-2959]{Jose A. Mu\~noz}
\email{Jose.A.Munoz@uv.es} 
\affiliation{Departamento de Astronom\'{\i}a y Astrof\'{\i}sica, Universidad de Valencia, E-46100 Burjassot, Valencia, Spain}
\affiliation{Observatorio Astron\'omico, Universidad de Valencia, E-46980 Paterna, Valencia, Spain}

\author[0000-0001-9087-1245]{Felipe Murgas}
\email{fmurgas@iac.es } 
\affiliation{Instituto de Astrof\'isica de Canarias (IAC), Calle V\'ia L\'actea s/n, 38200, La Laguna, Tenerife, Spain}
\affiliation{Departamento de Astrof\'isica, Universidad de La Laguna (ULL), E-38206 La Laguna, Tenerife, Spain}

\author[0000-0003-0987-1593]{Enric Palle}
\email{epalle@iac.es} 
\affiliation{Instituto de Astrof\'isica de Canarias (IAC), Calle V\'ia L\'actea s/n, 38200, La Laguna, Tenerife, Spain}
\affiliation{Departamento de Astrof\'isica, Universidad de La Laguna (ULL), E-38206 La Laguna, Tenerife, Spain}

\author[0000-0003-0647-6133]{Ivan A. Strakhov}
\email{strakhov.ia15@physics.msu.ru}
\affiliation{Sternberg Astronomical Institute, M.V. Lomonosov Moscow State University, 13, Universitetskij pr., 119234, Moscow, Russia}

\author[0000-0001-8227-1020]{Richard P. Schwarz}
\email{rpschwarz@comcast.net} 
\affiliation{Center for Astrophysics \textbar \ Harvard \& Smithsonian, 60 Garden Street, Cambridge, MA 02138, USA}

\author[0000-0002-1836-3120]{Avi Shporer}
\email{shporer@mit.edu} 
\affiliation{Department of Physics and Kavli Institute for Astrophysics and Space Research, Massachusetts Institute of Technology, Cambridge, MA 02139, USA}

\author{Gregor Srdoc}
\email{gregorsrdoc@gmail.com} 
\affil{Kotizarovci Observatory, Sarsoni 90, 51216 Viskovo, Croatia}

\author[0000-0003-2163-1437]{Chris Stockdale}
\email{thestockdalefamily@bigpond.com} 
\affiliation{Hazelwood Observatory, Australia}

\author[0000-0002-2146-3894]{Neil B.\ Thomas}
\email{nthomas819@aol.com} 
\affiliation{Department of Astronautics (ret), United States Air Force Academy, CO 80840, USA}

\author[0000-0003-2127-8952]{Francis P. Wilkin}
\email{wilkinf@union.edu} 
\affiliation{Department of Physics and Astronomy, Union College, 807 Union St., Schenectady, NY 12308, USA}

\begin{abstract}

    Warm Jupiters, with orbital periods of $10$–$200~\rm{days}$ and radii exceeding $8~R_{\oplus}$, are a relatively understudied class of exoplanets occupying the parameter space between hot Jupiters and more widely separated, colder Jupiter analogs. In this work, we report the detection of a multi-planet warm Jupiter system around TOI-3850 (TIC-143008050), a moderately active, near-solar metallicity G0 dwarf star observed by TESS in Sectors 15, 21, 41, 48 and 75. Initially, a single candidate planet was discovered by TESS, displaying transit timing variations (TTVs) with an amplitude of $\sim 1~\rm hr$ and a super-period of $513~\rm days$. Through a combination of transit photometry, radial velocity observations with MAROON-X, and TTV modeling, we identify two planets: TOI-3850 b $(P_b=14.484\pm0.002~\mathrm{days},~ M_b =112\pm20~M_{\oplus},~e_b = 0.018\pm0.008, R_b = 12.07\pm0.09~R_{\oplus}, ~T_{\rm{eq}}=841\pm10~\rm{K})$, a transiting warm Jupiter, and TOI-3850 c $(P_c=29.85\pm0.01~\mathrm{days},~ M_c =90\pm15~M_{\oplus},~e_c < 0.015, ~T_{\rm{eq}}=661\pm7~\rm{K})$, a non-transiting, Saturn-mass companion. The two planets lie wide of the 2:1 mean motion resonance $(P_c/P_b \approx 2.06)$, consistent with a formation history involving disk-driven migration. $N$-body integrations indicate that TOI-3850 c may begin to transit on decadal timescales, while TOI-3850 b remains a promising target for follow-up atmospheric characterization.

\end{abstract}


\section{Introduction}

Giant planets on short-period orbits provide crucial insights into planet formation and migration theories. Hot Jupiters $(R_p>8~R_{\oplus}, ~P < 10~\rm{days})$ are believed to form either close to their host stars or beyond the snowline and migrate inwards through disk-driven or high-eccentricity migration \citep{fortney2021, heller2019, batygin2016}. Although there exists a small population of exceptions (e.g., WASP-47b; \citealt{becker2015}), their overall lack of nearby companions suggests that many hot Jupiters undergo dynamically disruptive migration pathways that destabilize close companions \citep{mustill2015}.

Warm Jupiters occupy an intermediate regime between short-period hot Jupiters and more distant Jupiter analogs \citep{huang2016, dawson2018}. With orbital periods from $10$ to $200 ~ \rm{days}$ and radii greater than $8~R_{\oplus}$, warm Jupiters receive lower stellar irradiation than hot Jupiters while remaining accessible to transit, radial velocity (RV) and atmospheric follow-up. Consequently, their atmospheres may support processes distinct from those of highly irradiated hot Jupiters, making them interesting targets for atmospheric characterization \citep{fortney2021, rauscher2017}.

The origins of warm Jupiters remain uncertain. In certain models, warm Jupiters are interpreted as progenitors of hot Jupiters \citep{gupta2024, petrovich2016} while other theories suggest separate evolutionary histories \citep{anderson2020}. Proposed formation mechanisms include disk-driven migration where the planet exchanges angular momentum with the protoplanetary disk and spirals inwards \citep{anderson2017}; high-eccentricity migration in which gravitational perturbations from planetary or stellar companions cause an extreme increase in orbital eccentricity \citep{dong2013}; and in situ formation where the planet forms at its current location \citep{huang2016}. Each channel predicts different distributions of orbital eccentricity, mutual inclinations and companion architectures. Since warm Jupiters remain relatively understudied, a statistically meaningful sample of confirmed warm Jupiters is needed to test these different evolutionary pathways \citep{daoust2025}.

TOI-3850.01, with an orbital period of $14.5~\rm{days}$, is a candidate warm Jupiter orbiting its G0 dwarf star TOI-3850. TOI-3850.01 shows large deviations from a linear transit ephemeris, or transit timing variations (TTVs), which suggest the presence of at least one non-transiting companion planet. The candidate planet has also undergone an extensive follow-up campaign through the TESS Follow-up Observing Program, ruling out common false-positive signals such as stellar-mass companions.

In this work, we report the discovery of a pair of warm, Saturn-mass planets around TOI-3850 by combining the observed TTVs with RVs taken with the MAROON-X spectrograph. In \S \ref{sec:observations}, we outline the photometry, spectroscopy and high-resolution imaging used to confirm the planetary nature of the candidate planets. In \S \ref{chap:stellar_params}, we characterize the host star TOI-3850 while in \S \ref{sec:planet}, we determine the planetary parameters of the system. In \S \ref{sec:discussion}, we place the planets in broader context and provide an overview of the dynamics of the system before making concluding remarks in \S \ref{sec:conclusion}.

\section{Observations} \label{sec:observations}

\subsection{Transiting Exoplanet Survey Satellite}
\setcounter{footnote}{0}
\label{tess_obs}

NASA's Transiting Exoplanet Survey Satellite (TESS) is an all-sky transit survey designed to detect exoplanets around bright, nearby stars \citep{winn2025}. Launched in 2018, TESS has identified more than 8000\footnote{As of July 28, 2026, the NASA Exoplanet Archive lists 8064 TESS project candidates. \url{https://exoplanetarchive.ipac.caltech.edu/}.} candidate exoplanets over the course of its mission. With a field of view of $24^{\circ}\times96^\circ$ and a 13.7 day elliptical orbit around the Earth, TESS splits the sky into multiple Sectors, which are each observed for approximately 27 days. As a result, planets with orbital periods greater than this baseline are difficult to detect.

Each of TESS' four cameras takes an image every 2 seconds. In order to handle this large data volume, TESS co-adds these exposures on board the spacecraft to produce longer cadence data. In every Sector, 20,000 high-interest stars are designated as ``postage stamps" \citep{guerrero2021}. Targets selected as postage stamps are processed by the Science Processing Operations Center (SPOC) pipeline, which produces light curves at 2-minute cadence \citep{jenkinsSPOC2016}. In addition to these postage stamp targets, SPOC also extracts light curves from the full frame images (FFIs) for a large portion of stars in each Sector. The Quick Look Pipeline (QLP) independently produces light curves from FFIs for all stars brighter than $T_{\rm mag}=13.5$. The FFIs cover a $24^\circ\times 24^{\circ}$ FOV per camera and were initially produced at a cadence of $1800~\rm{s}$, later reduced to $600~\rm{s}$ and most recently, to $200~\rm{s}$.

\subsubsection{Initial TESS Detection}

A periodic signal in the light curve of TIC-143008050 was first identified as a threshold-crossing event in Sector 21. Since TIC-143008050 has a TESS magnitude of $T_{\rm mag} = 12.47$, the signal was vetted in an independent search for planets around faint stars \citep{kunimoto2022}, which extends beyond the standard QLP limit of $T_{\rm mag} < 10.5$ \citep{guerrero2021}. After completing data validation and vetting, the signal was promoted to TOI-3850.01 on 2021 June 23. The search uncovered a candidate planet with orbital period $P=14.48329\pm0.00005~ \rm days$, transit epoch $T_0 = 2459614.177 \pm 0.002~\rm BJD$ and radius $R = 13 \pm 2~\rm R_{\oplus}$.

TOI-3850 has been observed in Sectors 15, 21, 41, 48, and 75. All five Sectors include QLP light curves extracted from FFIs with cadences of $1800~\rm s$ in Sectors 15 and 21, $600~\rm s$ in Sectors 41 and 48, and $200~\rm s$ in Sector 75. Additionally, TOI-3850 was selected as a postage stamp and was processed by SPOC in Sectors 48 and 75, where 2-minute cadence light curves are available.

\subsubsection{This Work}

In this study, we use the publicly available FFI light curves for Sectors 15, 21, and 41 \citep{2020RNAAS...4..201C}. For Sectors 48 and 75, we instead use the SPOC 2-minute cadence light curves since their higher time resolution provides improved sampling of both the transit shape and mid-transit times. Once extracted, the light curves were cleaned by removing points flagged as poor quality and normalized by their median flux in each Sector. We obtained all publicly available TESS light curve products from the Mikulski Archive for Space Telescopes (MAST) using the \texttt{lightkurve} Python package \citep{lightkurve2018}.

To assess whether the transit signal occurs on the target star, we examined the individual TESS pixels. Due to the large TESS pixel scale, flux from background eclipsing binary stars (BEBs) can contaminate the photometric aperture, resulting in transit-like signals \citep{sullivan2015}. Difference imaging, which computes the difference between average in- and out-of-transit pixel values, can be used to identify BEBs \citep{bryson2013}. Because the centroid of the difference image shows the location of the transit source, these images can be used to find the true location of the periodic signal. 

\begin{figure}[htbp]
    \centering
    \includegraphics[width=0.99\linewidth]{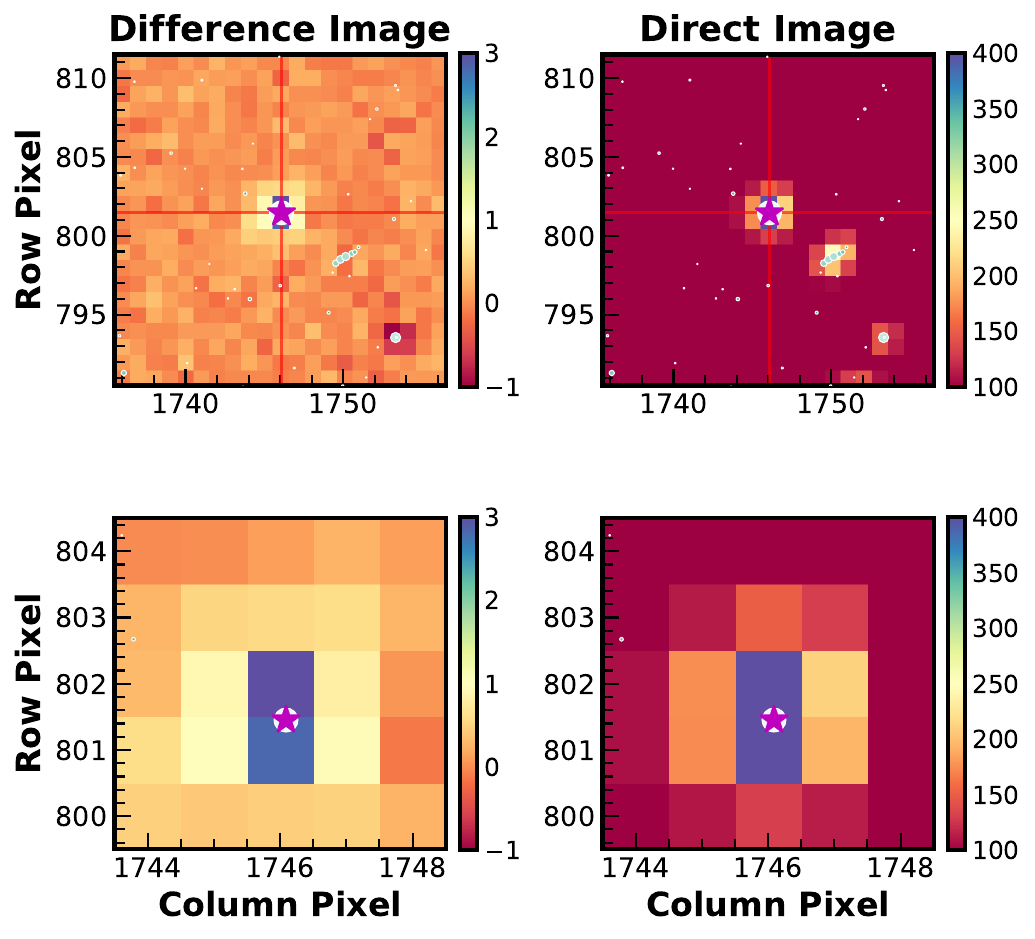}
    \caption{\textit{Left:} Difference image of TOI-3850.01. The top panel was produced using a $20\times20$ pixel grid of Sector 48 FFIs. The bottom panel zooms in on the central $5\times5$ pixels. \textit{Right:} Average of out-of-transit images, representing a direct image of the sky near TOI-3850. All colour bars show the flux in units of electrons per second. Nearby stars with $\Delta T_{\rm mag} \leq 10$ are depicted as white circles with sizes scaled by their flux. The target star TOI-3850 is shown as a purple central star. Due to the similarities between the difference and direct image, this analysis suggests the transit source is located on TOI-3850. }
    \label{fig:diffimage}
\end{figure}

Here, we build a difference image of TOI-3850 using the Python package \texttt{transit-diffImage}\footnote{\url{https://github.com/stevepur/transit-diffImage/tree/os-independence}} (see Figure~\ref{fig:diffimage}). This analysis suggests that the transit source is co-located on TOI-3850.

\subsection{Ground-Based Photometry}
\label{sec:ground_phot}

We obtained ground-based follow-up photometric observations through the TESS Follow-up Observing Program (TFOP; \citealt{exofop5}). TFOP concentrates on obtaining ground-based observations needed to validate TESS planet candidates. All available transit light curves can be found on the \texttt{ExoFOP} website.\footnote{\url{https://exofop.ipac.caltech.edu/tess/target.php?id=143008050}} We summarize our observations in Table $\ref{tab:observations}$.

Due to the large pixel scale of TESS, flux from nearby stars can contaminate the photometric aperture. Ground-based observations can rule out false positives due to BEBs at smaller angular separations by providing superior angular resolution than TESS. These observations can confirm the transit occurs on target. In addition, ground-based photometric observations are typically taken at higher cadence than TESS observations, allowing us to obtain precise constraints on the transit ephemeris. The resulting mid-transit times reveal large TTVs with an amplitude of $\sim1~\mathrm{hr}$ and a super-period\footnote{For TTVs, the super-period is the period of the TTV signal. It is given by $P_{\rm{super}} = |J/P_{\rm{outer}}-(J-1)/P_{\rm{inner}}|^{-1}$, where $J$ is a positive integer \citep{hadden2016}.} of $513~\mathrm{days}$. These TTVs provide evidence for the presence of a second, non-transiting companion planet (see \S\ref{sec:planet}).

\subsubsection{Las Cumbres Observatory}

Transits of TOI-3850.01 were observed with telescopes from the Las Cumbres Observatory Global Telescope network (LCOGT; \citealt{brown2013}). Observations taken by the $0.35~\rm{m}$ telescopes at the Teide (LCO-Teid) and Haleakala (LCO-HAL) observatories used a QHY600 camera with a pixel scale of $0.73''\,\mathrm{pixel}^{-1}$. Observations taken by the $0.35~\rm m$ telescope at the McDonald Observatory (LCO-McD) and the $0.4~\rm m$ at LCO-Teid used a QHY600 camera with a pixel scale of $0.57''\,\mathrm{pixel}^{-1}$. Observations taken by the $2~\rm m$ telescope at LCO-HAL used a MUSCAT3 camera with a pixel scale of $0.265''\,\mathrm{pixel}^{-1}$. All images were calibrated with the LCOGT \texttt{BANZAI} pipeline \citep{mccully2018} and converted into light curves using \texttt{AstroImageJ} \citep{collins2017}.

A full transit of TOI-3850.01 was observed on 2023 April 14 from LCO-McD in the \textit{ip} band using an uncontaminated $8.8''$ aperture. The ingress of this event was independently captured by the $0.4~\rm m$ telescope at LCO-Teid in the \textit{ip} band with an uncontaminated $5.7''$ aperture.

A full transit was observed on 2024 April 10 with the $2~\rm m$ telescope at LCO-HAL in the \textit{gp}, \textit{rp}, \textit{ip} and \textit{zs} bands. The observation used $6.4''$ target apertures and found that the transit depth is consistent across the bands. On 2024 May 9, the same telescope and bands were used to observe egress with target apertures ranging between $6.4''$ and $6.6''$.

Additional full transits were observed on 2025 February 8, 2025 April 21, 2025 June 18, 2026 April 4 and 2026 May 17 with the $0.35~\rm m$ at LCO-HAL, the $0.35~\rm m$ at LCO-Teid and the $0.4~\rm m$ at LCO-Teid. All five observations were conducted in the \textit{gp} band.

\subsubsection{Fred Lawrence Whipple Observatory}

Multiple transits of TOI-3850.01 were observed using the $1.2~\rm m$ telescope at the Fred Lawrence Whipple Observatory (FLWO). The telescope is equipped with the KeplerCam and has a pixel scale of $0.672''\,\mathrm{pixel}^{-1}$. Full transits were observed on 2022 January 5 and 2024 April 10. Egress was observed on 2022 May 16 and 2023 April 14, and ingress on 2025 February 8. All observations were conducted in the \textit{ip} band.

\subsubsection{Observatori Astronòmic de la Universitat de València}

Two partial transits were observed with the $0.5~\rm m$ T50 telescope at the Observatori Astronòmic de la Universitat de València (OAUV). Both observations were conducted in the \textit{R} band and utilized a pixel scale of $0.54''\,\mathrm{pixel}^{-1}$. Using a FLI Proline camera, egress was observed on 2022 January 19. Ingress was observed on 2025 January 24 with a FLI Kepler camera.

\subsubsection{Meade LX850}

Ingress was observed on 2023 March 15 using a Meade LX850 (MLX) telescope and a QHY268m camera with a pixel scale of $0.3''\,\mathrm{pixel}^{-1}$. The observations were conducted in the \textit{Red} band and used an uncontaminated $5.4''$ aperture. 

\subsubsection{Acton-Sky-Portal}

We obtained two transit observations with the $0.36~\rm m$ telescope at the private Acton Sky Portal (ASP) observatory using a SBIG A4710 camera. The camera has a pixel scale of $1''\,\mathrm{pixel}^{-1}$ and both observations were taken in the \textit{rp} band. A full transit was observed on 2023 April 14 while ingress was observed on 2023 December 16. The transit on 2023 April 14 was also observed with FLWO, LCO-Teid and LCO-McD.

\subsubsection{Lookout Observatory}

A full transit was observed with the $0.27~\rm m$ telescope at the Lookout Observatory (LKO) using an ASI 2600 camera with a pixel scale of $2.92''\,\mathrm{pixel}^{-1}$. The observations were obtained with a one-shot color (OSC) detector, corresponding to a broad bandpass. 

\subsubsection{Canis Major Observatory}

We observed a full transit with the $0.4~\rm m$ telescope at the Canis Major Observatory (CMO) using a SBIG STXL camera with a pixel scale of $0.73''\,\mathrm{pixel}^{-1}$. The transit was observed in the \textit{ic} band.

\subsubsection{The Large Array Survey Telescope}

A full transit was observed with the $0.4~\rm m$ Large Array Survey Telescope (LAST) using a QHY600M camera with a pixel scale of $1.25''\,\mathrm{pixel}^{-1}$. These observations were taken without the use of a filter.

\begin{table*}[htbp]
\centering
\caption{Summary of ground-based photometric observations of TOI-3850.01.}
\label{tab:observations}
\vspace{0.5em}

\begin{tabular*}{0.9\textwidth}{@{\extracolsep{\fill}}lccccc@{}}
\toprule
Telescope & Date & Aperture & Filter & Notes & Included in Fits \\
& (UTC) & ($''$) & & & (Y/N) \\
\midrule
FLWO & 2022-01-05 & 5.4 & \textit{ip} & Full transit & N \\
OAUV & 2022-01-19 & 6.5 & \textit{R} & Egress & N \\
FLWO & 2022-05-16 & 5.4 & \textit{ip} & Egress & Y \\
MLX & 2023-03-15 & 5.4 & \textit{Red} & Ingress & N \\
FLWO & 2023-04-14 & 5.4 & \textit{ip} & Egress & Y \\
LCO-McD & 2023-04-14 & 8.8 & \textit{ip} & Full transit & Y \\
ASP & 2023-04-14 & 6.0 & \textit{rp} & Full transit & Y \\
LCO-Teid & 2023-04-13 & 5.7 & \textit{ip} & Ingress & Y \\
ASP & 2023-12-16 & 6.0 & \textit{rp} & Ingress & N \\
LCO-HAL & 2024-04-10 & 6.4 & \textit{gp, rp, ip, zs} & Full transit & Y \\
LCO-HAL & 2024-05-09 & 6.6 & \textit{gp, rp, ip, zs} & Egress & Y \\
FLWO & 2024-04-10 & 5.4 & \textit{ip} & Full transit & Y \\
LKO & 2024-02-12 & 20.4 & \textit{OSC} & Full transit & N \\
OAUV & 2025-01-24 & 6.5 & \textit{R} & Ingress & Y \\
LCO-HAL & 2025-02-08 & 7.3 & \textit{gp} & Full transit & Y \\
FLWO & 2025-02-08 & 5.4 & \textit{ip} & Ingress & Y \\
LAST & 2025-03-23 & 8.8 & None & Full transit & N \\
LCO-Teid & 2025-04-21 & 5.4 & \textit{gp} & Full transit & Y \\
LCO-Teid & 2025-06-18 & 5.8 & \textit{gp} & Full transit & Y \\
CMO & 2026-03-20 & 8.0 & \textit{ic} & Full transit & Y \\
LCO-HAL & 2026-04-04 & 5.9 & \textit{gp} & Full transit & Y \\
LCO-Teid & 2026-05-17 & 8.8 & \textit{gp} & Full transit & Y \\
\bottomrule
\end{tabular*}
\end{table*}

\subsection{High-Resolution Imaging}
\label{sec:high_res}

In order to search for nearby contaminants and companion stars, we obtained a high-resolution image from the Caucasian Observatory of the Sternberg Astronomical Institute (SAI) of Lomonosov Moscow State University \citep{strakhov2023}. High-resolution imaging can potentially detect unresolved BEBs that may be causing false positive signals. Similarly, light from nearby stars can dilute the transit depth, causing a bias towards smaller planet radii.

\begin{figure}[htbp]
    \centering
    \includegraphics[width=0.99\linewidth]{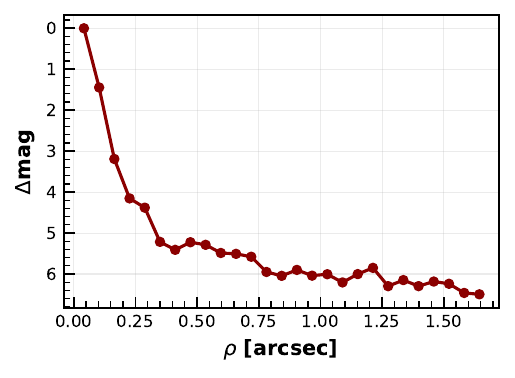}
    \caption{High-resolution contrast curve for TOI-3850, derived from speckle observations in the $I$-band with the SAI-2.5 m telescope. The curve shows the limiting difference in apparent magnitude between the target and a detectable companion as a function of angular distance from TOI-3850. The increasing contrast toward larger separations indicates sensitivity to fainter companions farther from the target. No nearby companion stars were detected within the sensitivity of the instrument.}
    \label{fig:speckle}
\end{figure}

TOI-3850 was observed on 2023 January 18 with a speckle polarimeter on the $2.5~\rm m$ telescope in the \textit{I} band. Observations were made with an estimated PSF of $0.083''$ and achieved a contrast of $\Delta T_{\rm{mag}} = 6$ at a separation of $1''$ from the target star (see Figure~\ref{fig:speckle}). No contaminants were detected within $\sim1.5''$ of the target.

\subsection{High-Resolution Spectroscopy}
\label{sec:spectroscopy}

Spectroscopic observations allow us to characterize the host star from its spectral lines. RV measurements derived from spectroscopic observations can rule out false positives caused by eclipsing binaries by placing upper limits on the mass of the companion(s). Precision RV data can constrain planet mass and lead to the confirmation of planetary candidates.

\subsubsection{TRES}

We obtained two reconnaissance spectra of TOI-3850 with the Tillinghast Reflector Echelle Spectrograph (TRES) on 2022 January 8 and 2022 January 31. TRES is mounted on the $1.5~\rm m$ Tillinghast optical spectroscopic telescope at the FLWO. It provides wavelength coverage from $3850$–$9096$ \AA\ at a resolving power of $R \sim 44,000$ across 51 orders \citep{law2017}. The two spectra placed an upper limit on the RV semi-amplitude of $K<150~\rm{m~s^{-1}}$, ruling out the presence of a stellar-mass companion. The spectra also suggest that TOI-3850 is a moderately rotating G dwarf and is therefore amenable to high-precision RV follow-up observations.

\subsubsection{MAROON-X}

We obtained 14 spectra of TOI-3850 (see Table~\ref{tab:rv_measurements}) between December 2025 and April 2026 with the MAROON-X spectrograph located on the $8.1~\rm m$ Gemini North telescope atop Maunakea \citep{seifahrt2018}. An exposure time of $1000~\rm{s}$ was used for each observation. MAROON-X is a fiber-fed echelle spectrograph with a high resolving power ($R\sim80,000$) across its 56 orders. Its blue arm provides wavelength coverage from $500-670~\rm{nm}$ while the red arm covers wavelengths $650-900~\rm{nm}$. MAROON-X has a long-term instrumental stability better than $0.7~\rm{m~s^{-1}}$ and has been used to measure the masses of several planets \citep{winters2022, trifonov2021}.

The spectra were reduced using a custom Python 3 data reduction pipeline. The RVs were calculated using the \texttt{SERVAL} pipeline \citep{zechmeister2018}, which employs a least-squares fitting algorithm that matches spectral templates to each order. To mitigate the effects of Earth's atmosphere, \texttt{SERVAL} masks telluric features deeper than 5\%. 

We excluded one observation from the RV analysis since it was a significant outlier relative to the remaining measurements and lies 5$\sigma$ away from the posterior median solution presented in \S \ref{sec:joint}. Because the spectra have consistently higher signal-to-noise ratios in the red arm, we restrict our RVs to the red arm in the following analysis. The 13 remaining RVs have an average uncertainty of 10.4$~\rm{m~s^{-1}}$ and span a baseline of 124 days.

\section{Stellar Parameters} \label{chap:stellar_params}

\subsection{TESS Input Catalog}
\label{sec:TIC}

Our initial estimates of the stellar parameters of TOI-3850 (TIC-143008050) were obtained from the TESS Input Catalog (TIC; \citealt{stassun2019}), which provides estimates of stellar properties for all TESS targets. To do this, the TIC compiles astronomical data from multiple surveys, such as Gaia and 2MASS. TICv8.2 lists TOI-3850 with the following properties: $R_{\star} =1.02\pm0.05~R_{\odot}$, $M_{\star}=1.12\pm0.14~ M_{\odot}$, $\log(g) = 4.47\pm0.08$ (cgs) and $T_{\rm eff} = 6039\pm125~\rm K$. These parameters are consistent with TOI-3850 being a G0 main-sequence star. We summarize the stellar properties of TOI-3850 in Table~\ref{tab:stellar_params}.

\begin{table*}[htbp]
\centering
\caption{Stellar parameters for TOI-3850 (TIC-143008050).}
\label{tab:stellar_params}
\vspace{0.5em}
\renewcommand{\arraystretch}{1.15}
\begin{tabular*}{0.95\textwidth}{@{\extracolsep{\fill}}llll@{}}
\toprule
Parameter & Value & Description & Source \\
\midrule

\multicolumn{4}{c}{\textit{TIC Parameters}} \\
\midrule
ID & 143008050 & TESS Input Catalog ID & TICv8.2 \\
$T_{\rm eff}$ & $6039 \pm 125$ & Effective temperature (K) & TICv8.2 \\
$\log g$ & $4.47 \pm 0.08$ & Surface gravity (cgs) & TICv8.2 \\
$R_\star$ & $1.02 \pm 0.05$ & Stellar radius ($R_\odot$) & TICv8.2 \\
$M_\star$ & $1.12 \pm 0.14$ & Stellar mass ($M_\odot$) & TICv8.2 \\

\midrule
\multicolumn{4}{c}{\textit{Spectroscopic Parameters}} \\
\midrule
$T_{\rm eff}$ & $5906 \pm 50$ & Effective temperature (K) & TRES/SPC \#1$^{a}$ \\
$\log g$ & $4.459 \pm 0.10$ & Surface gravity (cgs) & TRES/SPC \#1 \\
$[\mathrm{Fe/H}]$ & $0.102 \pm 0.08$ & Metallicity & TRES/SPC \#1 \\
$v \sin i$ & $7.0 \pm 0.5$ & Projected rotational velocity ($\mathrm{km\,s^{-1}}$) & TRES/SPC \#1 \\

\midrule
$T_{\rm eff}$ & $6090 \pm 62$ & Effective temperature (K) & TRES/SPC \#2$^{b}$ \\
$\log g$ & $4.48 \pm 0.11$ & Surface gravity (cgs) & TRES/SPC \#2 \\
$[\mathrm{Fe/H}]$ & $0.266 \pm 0.08$ & Metallicity & TRES/SPC \#2 \\
$v \sin i$ & $6.98 \pm 0.5$ & Projected rotational velocity ($\mathrm{km\,s^{-1}}$) & TRES/SPC \#2 \\

\midrule
\multicolumn{4}{c}{\textit{SED Parameters}} \\
\midrule
$M_\star$ & $1.14\pm0.07$ & Stellar mass ($M_\odot$) & SED analysis \\
$R_\star$ & $1.010 \pm 0.036$ & Stellar radius ($R_\odot$) & SED analysis \\
$F_{\rm{bol}}$ & $(1.711 \pm 0.040)\times10^{-10}$ & Bolometric flux (erg~s$^{-1}$~cm$^{-2}$) & SED analysis \\
$L_{\rm{bol}}$ & $1.184\pm0.029$ & Bolometric luminosity ($L_{\odot}$) & SED analysis \\

\midrule
\multicolumn{4}{c}{\textit{Isochrone-derived Parameters}} \\
\midrule
$M_\star$ & $1.10\pm0.02$ & Stellar mass ($M_\odot$) & \texttt{isochrones} fit \\
$R_\star$ & $1.03 \pm 0.01$ & Stellar radius ($R_\odot$) & \texttt{isochrones} fit \\
$T_{\rm eff}$ & $5956\pm58$ & Effective temperature (K) & \texttt{isochrones} fit \\
$\log g$ & $4.45 \pm 0.01$ & Surface gravity (cgs) & \texttt{isochrones} fit \\
$[\mathrm{Fe/H}]$ & $0.10\pm0.05$ & Metallicity & \texttt{isochrones} fit \\
$\tau_{\star}$ & $0.89^{+1.01}_{-0.63}$ & Stellar age (Gyr) & \texttt{isochrones} fit \\
$d_{\star}$ & $482\pm5$ & Distance (pc) & \texttt{isochrones} fit \\

\bottomrule
\end{tabular*}

\vspace{0.5em}
\begin{minipage}{0.95\textwidth}
\footnotesize
$^{a}$ Derived from the TRES spectrum taken on 2022 January 8.\\
$^{b}$ Derived from the TRES spectrum taken on 2022 January 31.
\end{minipage}

\end{table*}

\subsection{Spectroscopic Parameters}
\label{sec:spec_params}

To estimate the spectroscopic parameters from the two TRES spectra, we utilized the results from the Stellar Parameter Classification tool (SPC; \citealt{buchhave2012}). The SPC derives stellar properties by cross-correlating the observed stellar spectra to a bank of synthetic templates. All models have varying values of effective temperature $T_{\rm eff}$, stellar metallicity $[\rm Fe/H]$, surface gravity $\log(g)$, and projected rotational velocity $v\sin(i)$.  For the spectrum taken on 2022 January 8, the SPC estimates $T_{\rm eff} = 5906\pm50~\rm K$, $\log(g) = 4.459 \pm 0.1$ (cgs), $[\rm Fe/H] = 0.102\pm0.08$, and $v\sin(i) = 7.0 \pm 0.5~\rm km~s^{-1}$. For the second spectrum taken on 2022 January 31, the SPC estimates  $T_{\rm eff} = 6090\pm62~\rm K$, $\log(g) = 4.483 \pm 0.109$ (cgs), $[\rm Fe/H] = 0.266\pm0.08$, and $v\sin(i) = 6.98 \pm 0.5~\rm km~s^{-1}$. Similar to TICv8.2, these spectroscopic parameters suggest TOI-3850 is a G0 main-sequence star. Furthermore, the moderate projected rotational velocity of $v\sin(i) \approx 7~\rm km~s^{-1}$ indicates that TOI-3850 is amenable to follow-up precision RVs.

\subsection{SED Analysis}

\begin{figure*}[!ht]
    \centering
    \includegraphics[width=0.75\linewidth,trim=80 70 50 50,clip]{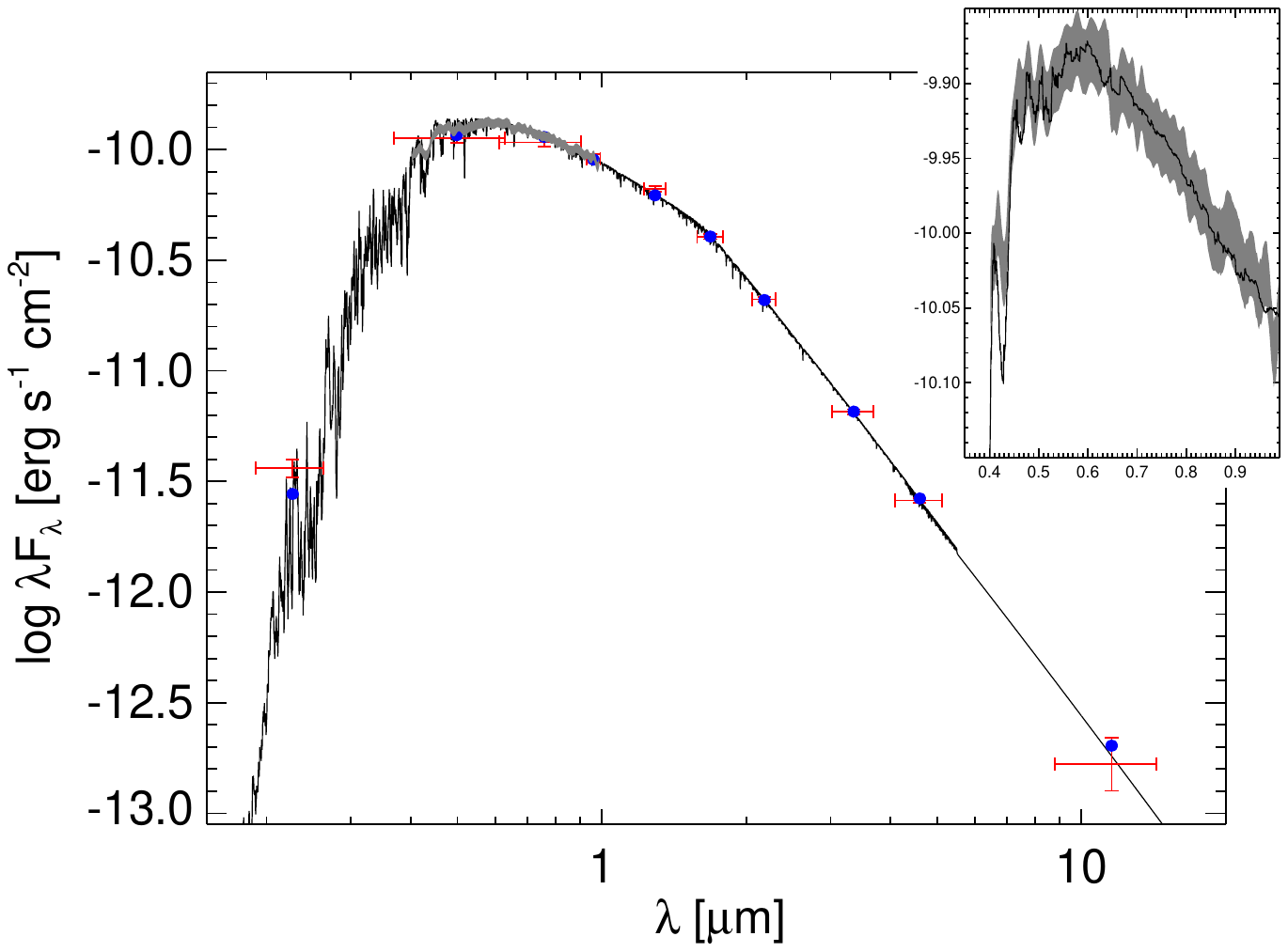}
\caption{Spectral energy distribution of TOI-3850. Red symbols represent the observed photometric measurements, where the horizontal bars represent the effective width of the passband. Blue symbols are the model fluxes from the best-fit PHOENIX atmosphere model (black). The gray swathe represents the {\it Gaia\/} spectrophotometry (also in the inset). \label{fig:sed}}
\end{figure*}

As an independent determination of the basic stellar parameters, we performed an analysis of the broadband spectral energy distribution (SED) of the star together with the {\it Gaia\/} DR3 parallax \citep[with no systematic offset applied; see, e.g.,][]{StassunTorres:2021}, in order to determine an empirical measurement of the stellar radius, following the procedures described in \citet{Stassun:2016,Stassun:2017,Stassun:2018}. We pulled the $JHK_S$ magnitudes from {\it 2MASS}, the $G_{\rm BP}, G_{\rm RP}$ magnitudes from {\it Gaia}, and the W1--W3 magnitudes from {\it WISE}. We also utilized the NUV magnitude from {\it GALEX} and the absolute flux-calibrated spectrophotometry from {\it Gaia}. Together, the photometry spans the full stellar SED over the wavelength range 0.2--10~$\mu$m (see Figure~\ref{fig:sed}).  

We performed a fit using PHOENIX stellar atmosphere models \citep{Husser:2013}, with the effective temperature ($T_{\rm eff}$), surface gravity ($\log g$), and metallicity ([Fe/H]) adopted from the spectroscopic analysis. The extinction, $A_V$, was limited to maximum line-of-sight value from the Galactic dust maps of \citet{Schlegel:1998}. The resulting fit (Figure~\ref{fig:sed}) has a reduced $\chi^2$ of 1.6, with a best-fit $A_V = 0.02 \pm 0.02$. Integrating the (unreddened) model SED gives the bolometric flux at Earth, $F_{\rm bol} = 1.711 \pm 0.040 \times 10^{-10}$ erg~s$^{-1}$~cm$^{-2}$. Taking the $F_{\rm bol}$ together with the {\it Gaia\/} parallax directly gives the bolometric luminosity, $L_{\rm bol} = 1.184 \pm 0.029$~L$_\odot$. The stellar radius follows from the Stefan-Boltzmann relation, giving $R_\star = 1.010 \pm 0.036$~R$_\odot$. In addition, we can estimate the stellar mass from the empirical relations of \citet{Torres:2010}, giving $M_\star = 1.14 \pm 0.07$~M$_\odot$. 

\subsection{Isochrones Analysis}
\label{sec:isochrone}

In this work, we perform isochrone fitting with the \texttt{isochrones} Python package \citep{morton2015} by comparing TOI-3850's position on a HR diagram to MIST evolutionary models \citep{choi2016}. As inputs, we use the observed photometric magnitudes ($J,H,K,W_1,W_2,W_3,G,G_{RP},G_{BP}$), parallax from Gaia DR3, and the error-weighted mean effective temperature, surface gravity and metallicity from the two TRES SPC fits. Adopting the convention from \citet{eastman2017} and \citet{kunimoto2023}, we inflate the uncertainties to 0.02 for Gaia magnitudes and 0.03 in WISE bands.

In our fits, we varied age, metallicity, distance, extinction, and the equivalent evolutionary phase (EEP), which is analogous to the star's stage in its overall stellar evolution. We set a uniform prior on the stellar metallicity between $-0.3 \leq [\rm Fe/H] \leq 0.3$ and used the default \texttt{isochrones} priors for the four remaining parameters. To sample the posterior distributions, we employed a Markov chain Monte Carlo (MCMC) approach using the \texttt{emcee} sampler \citep{foreman2013}, with $100$ walkers evolved over $40,000$ steps. The first $5000$ steps were discarded as burn-in steps and not used to calculate the final parameters.

From our \texttt{isochrones} fit, we find that TOI-3850 is consistent with a young $\left (\tau_{\star} = 0.89^{+1.01}_{-0.63}~\rm Gyr \right)$, near solar metallicity $\left ([\rm Fe/H] = 0.10\pm0.05 \right)$ G0 dwarf star with $T_{\rm eff} = 5956\pm 58~\rm K$, $M_{\star} = 1.10\pm0.02~ M_{\odot}$, and $R_{\star} = 1.03\pm0.01 ~ R_{\odot}$. For the remainder of this study, we adopt these parameters when estimates of the star's mass and radius are required.

\subsection{Stellar Activity}
\label{sec:activity}

\begin{figure*}[htbp]
    \centering
    \includegraphics[width=0.95\linewidth]{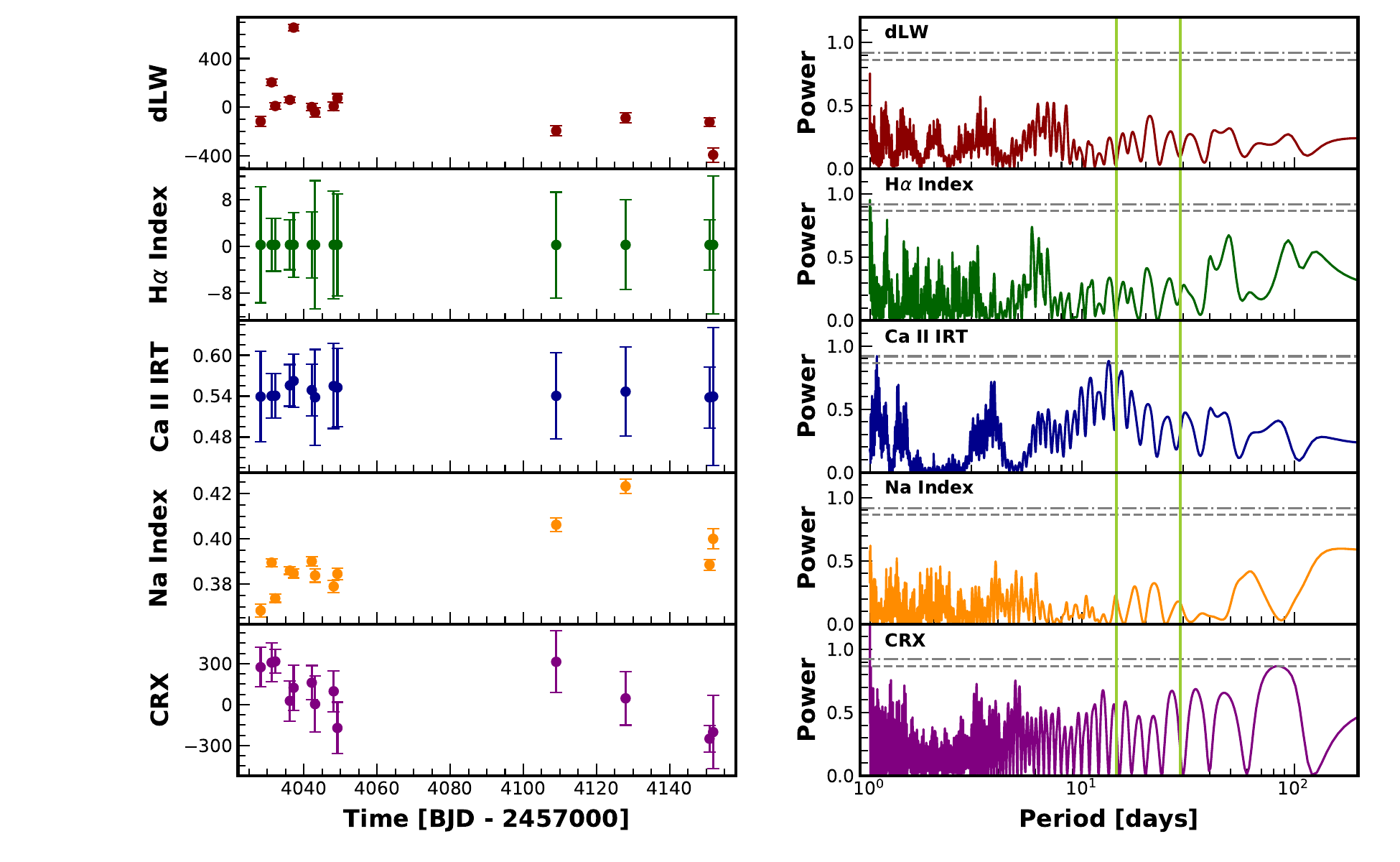}
    \caption{\textit{Left:} Time series of stellar activity indicators determined by the \texttt{SERVAL} reduction pipeline. Indicators include the differential line width ($\rm{dLW}$) and the $\rm{H}\alpha$, $\rm Ca~II~IRT$, $\rm Na$, and chromatic (CRX) indices. \textit{Right:} Generalized Lomb Scargle (GLS) periodogram of stellar activity indicators. The green vertical lines depict the orbital periods of the two planets. The grey horizontal lines indicate the 1\% (dashed) and 0.1\% (dashed-dotted) false alarm probabilities (FAP). We do not detect any significant stellar activity that may be mimicking planetary RVs.}
    \label{fig:indicators}
\end{figure*}

For each spectrum taken with MAROON-X, the \texttt{SERVAL} pipeline calculates five activity indicators: the differential line width ($\rm{dLW}$) and the $\rm{H}\alpha$, $\rm Ca~II~IRT$, $\rm Na$, and chromatic (CRX) indices. To check for periodic non-planetary signals in our observations, we compute the generalized Lomb-Scargle (GLS) periodograms for each of the five indicators (see Figure~\ref{fig:indicators}). The GLS periodogram is a statistical technique used to find periodic signals in unevenly spaced data points \citep{lomb1976, scargle1982, zechmeister2009}. In this work, we compute the GLS periodograms of the indicators using the \texttt{Astropy} implementation \citep{astropy2022}.

Figure~\ref{fig:indicators} shows the five activity indicators' time series and GLS periodograms obtained from the MAROON-X observations. In addition to the periodograms, we also plot the 1\% and 0.1\% false alarm probabilities (FAPs) as horizontal lines, and the orbital periods of the two candidate planets, TOI-3850.01 and its companion, as vertical lines. 

No significant peaks are observed above the FAP levels at the orbital periods of either TOI-3850.01 or the non-transiting companion in any of the indicators. This suggests the RV signals are driven by the two candidate planets rather than stellar activity. While there is a noticeable peak in the $\rm Ca~II~IRT$ periodogram around $\sim14~\rm days$, it lies below the 0.1\% FAP and may be driven by the large uncertainties in the  $\rm Ca~II~IRT$ indices. Similarly, the excess power around $1 ~ \rm{day}$ in the periodograms is likely associated with nightly observing cadence, rather than a physical stellar signal.

\begin{figure}[htbp]
    \centering
    \includegraphics[width=0.9\linewidth]{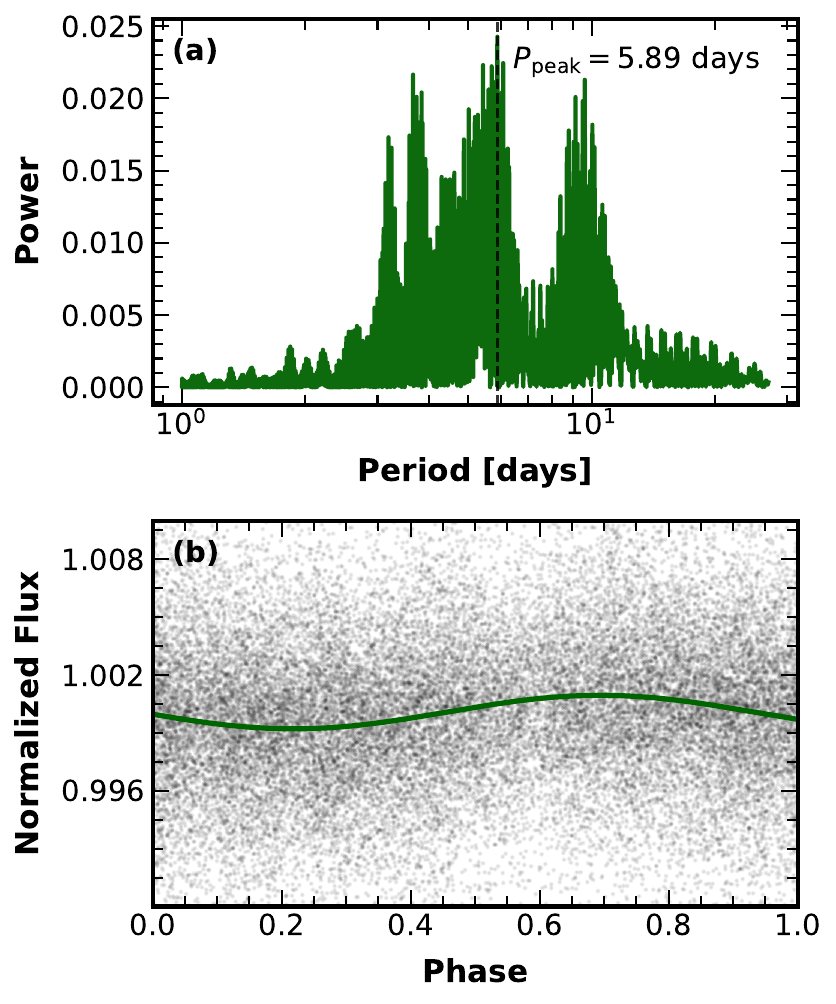}
    \caption{\textit{Panel (a).}  Generalized Lomb Scargle (GLS) periodogram of the out-of-transit TESS light curve in Sectors 15, 21, 41, 48, and 75. The peak periodic component at $P_{peak} = 5.89~\rm days$ suggests TOI-3850 has a rotation period $P \sim 6~\rm days$. Since this does not overlap with the orbital periods of TOI-3850.01 and its companion, there is no indication that stellar rotation is mimicking planetary RV signals. \textit{Panel (b).} Out-of-transit TESS light curves phase-folded at a period of $P_{\rm peak} = 5.89~\rm days$ and best-fit sine wave model in green. The sine wave model finds a best-fit amplitude of $870~\rm ppm$, corresponding to an order-of-magnitude RV jitter estimate of $\sigma \sim 10~\rm{m~s^{-1}}$.}
    \label{fig:rotation}
\end{figure}

Another potential cause for stellar activity is stellar rotation, which can mimic periodic planetary signals in RV analyses. To investigate this, we compute the GLS periodogram of the out-of-transit TESS light curve in Sectors 15, 21, 41, 48, and 75. We use the raw, non-detrended light curves to ensure we do not remove any periodic signals arising from stellar rotation. The periodogram is shown in Figure~\ref{fig:rotation} with a peak contribution at $P_{\rm peak} = 5.89~\rm days$, corresponding to a rotation period of $\sim 6~\rm days$. Since this rotation period does not overlap with the orbital periods of TOI-3850.01 and its companion, we find no evidence that stellar rotation is driving the observed RV variations. However, due to the 13.7 day period of the TESS orbit, it is difficult to detect rotation periods larger than $\sim13~\rm days$.

To estimate the expected activity-induced RV jitter (i.e., the RV variability arising from stellar activity rather than planetary motion), we phase-folded the out-of-transit TESS light curve at $P_{\rm peak} = 5.89~\rm days$ (see Figure~\ref{fig:rotation}). We then fit a simple sine wave model to the phase-folded data and estimate an amplitude of $870~\rm ppm$. Following \citet{saar1997}, we estimate the order-of-magnitude RV jitter as $\sigma \sim 10~\rm m~s^{-1}$, consistent with moderate levels of stellar activity.

\section{Planetary Parameters} \label{sec:planet}

\subsection{Transit-Only Fit}
\label{sec: transit}

To constrain the ephemeris of TOI-3850.01, we fit all available TESS light curves and select ground-based photometry. We use the FFI light curves for Sectors 15, 21, and 41 and the 2-minute SPOC light curves for Sectors 48 and 75. The TESS light curves were detrended with the \texttt{Wotan} package \citep{hippke2019} by using a window length of 1 day to remove long-term systematics while preserving the transit signal. We also include all transit observations from the LCO telescopes, FLWO observations on 2022-05-16, 2023-04-14, 2024-04-10 and 2025-02-08, ASP observations on 2023-04-14, CMO observations on 2026-03-20 and a single transit from OAUV on 2025-01-24 (see summary in Table~\ref{tab:observations}). This selection was made to ensure that the derived mid-transit times were constrained by observations with sufficient cadence, coverage and photometric precision.

The stellar radius and mass were set to the values obtained in our \texttt{isochrones} fit. Preliminary mid-transit times were estimated by individually fitting each transit with the \texttt{batman} Python package \citep{kreidberg2015}. We set all planetary parameters to those listed on the \texttt{ExoFOP} website and only let the mid-transit time vary. The mid-transit times were optimized via least-squares fitting with \texttt{LMFIT} \citep{newville2025}.

We parametrized our global transit model by the mid-transit times $t_n$, impact parameter $b$, and planet-to-star radius ratio $R_p/R_s$. A quadratic-limb darkening model was separately fit to the TESS, $ip$, $rp$, $gp$, and $R$ bands. We also individually fit for flux offsets and jitter terms to each Sector of TESS data and each ground-based observation to account for instrument-dependent systematics. For the global fit, we further assumed a circular orbit for TOI-3850.01 to reduce the number of free parameters and improve computational efficiency. It is expected that eccentricity will be better constrained by the RV data. 

We performed the global transit fit with \texttt{PyMC} \citep{abrilpla2023}. The transit model was constructed using the \texttt{exoplanet} Python package \citep{foreman2021}, which provides tools for modeling Keplerian orbits and their perturbations from companion planets. For each light curve, we integrated the transit model over the corresponding exposure time using an oversampling factor of three. We sampled the posterior distribution using four chains, each with 2000 tuning steps and 2000 draw steps. To assess convergence, we used the Gelman--Rubin statistic and found $\hat{R} < 1.01$ for all fitted parameters \citep{gelman1992}. Gaussian priors were placed on the mid-transit times, centered on the values from our \texttt{batman} fit. For all other free parameters, we set broad, uniform priors. 

\begin{figure*}[htbp]
    \centering
    \includegraphics[width=0.95\linewidth]{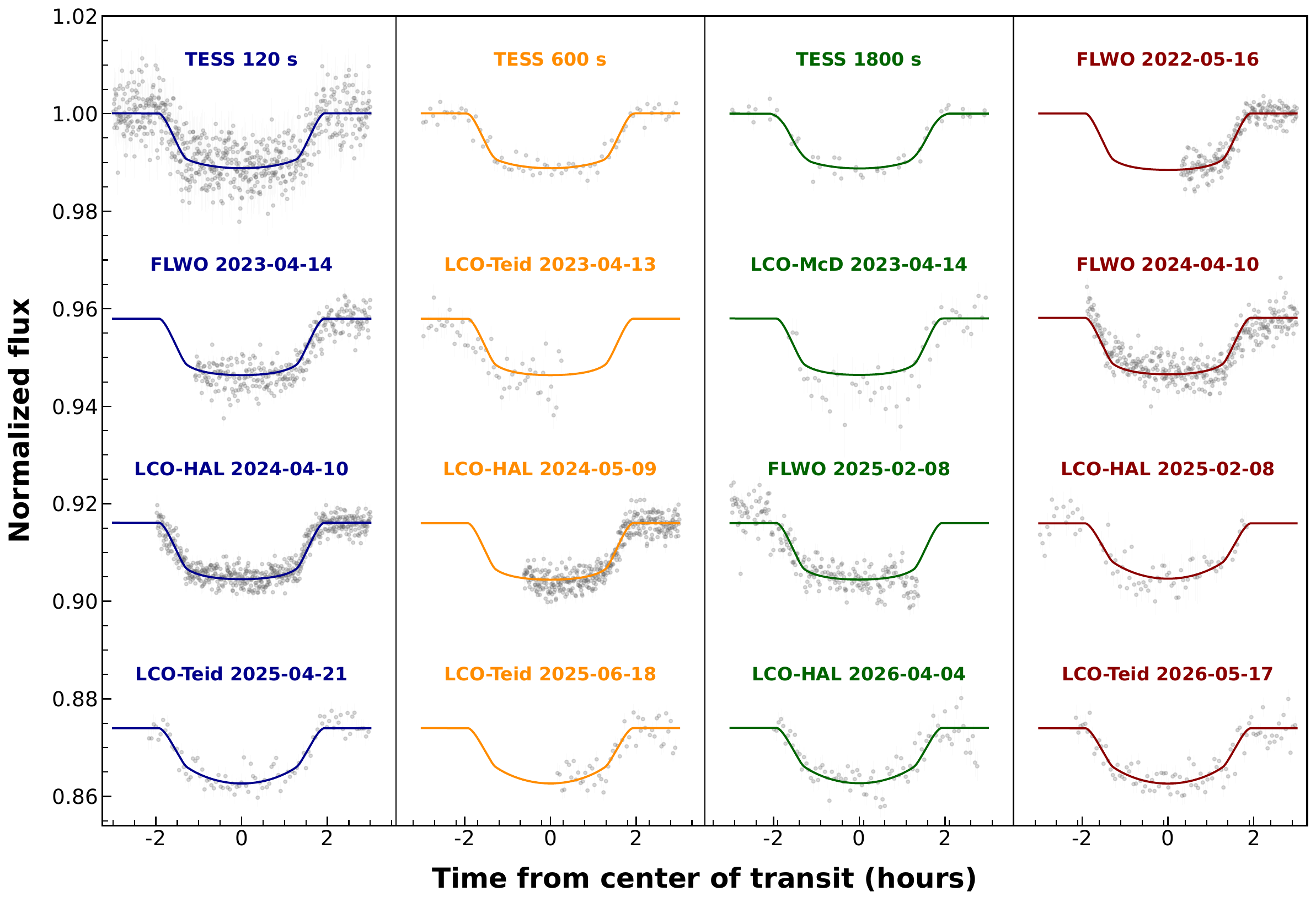}
    \caption{Median light curve models and transit data for all TESS, LCO, and FLWO observations used in the global transit fit. Each night of ground-based photometry is shown separately while the TESS observations have been phase-folded at the median orbital period. For clarity, we vertically shift the light curves by an arbitrary amount.}
    \label{fig:all_trans}
\end{figure*}

The individual median light curve models are shown in Figure~\ref{fig:all_trans} while the results of the global transit fit are given in Table~\ref{tab:planet_params}. From this analysis, we derive the mid-transit times, which are used to compute the TTVs (see Figure~\ref{fig:o-c} and Table~\ref{tab:transit_times}). We also find TOI-3850.01 has a radius of $R \approx 12~R_{\oplus}$, an orbital period of $P\approx14.484~\rm days$, and an orbital inclination of $i \approx 88.4^{\circ}$. The reported parameter values correspond to the median of the distributions and uncertainties given by the 16th and 84th percentiles.

\subsection{RV-Only Fit}
\label{sec:rv_only}

To verify that both planetary signals are present in the RV data, we first fit the MAROON-X observations independent of the observed TTVs. We considered three RV models: a null model with no planets, a one-planet model, and a two-planet model. The zero-planet (null) model consists of a constant flat line, representing a systematic RV offset while the one- and two-planet models consist of Keplerian signals. For each model, we also fit for an RV offset and jitter term added in quadrature to the observed MAROON-X uncertainties. In the two planet case, we test both scenarios where the companion is interior or exterior to TOI-3850.01.

In all models, we assumed circular orbits. For TOI-3850.01, we fixed the orbital period and transit epoch to the values inferred from the global transit fit. Due to the high amplitude and large super-period of the TTVs, we expect that the two planets have a period ratio near a first-order MMR \citep{agol2005, hadden2016}. To explore the possible architecture of the system, we tested four MMR configurations for the companion planet with period ratios of 2:1, 3:2, 4:3, and 5:4 relative to TOI-3850.01, along with the reciprocal configurations for an interior companion. 

By rearranging the definition of TTV super-period, the orbital periods of two planets near a first-order MMR are related by
\begin{equation}
    P_{\rm{inner}} = \frac{J-1}{\frac{J}{P_{\rm{outer}}}\pm\frac{1}{P_{\rm{super}}}}
    \label{eq:inner}
\end{equation}
and 
\begin{equation}
    P_{\rm{outer}} = \frac{J}{\frac{J-1}{P_{\rm{inner}}}\pm\frac{1}{P_{\rm{super}}}}
    \label{eq:outer}
\end{equation}
where $J>1$ is an integer. The two signs correspond to the two possible period branches located on either side of exact resonance. Using $P_{\rm{super}}=513~\rm{days}$, we tested the solutions corresponding to both of these branches.

For each MMR, we placed a Gaussian prior on the orbital period of the companion centered on the expected periods implied from Equations~(\ref{eq:inner}) and~(\ref{eq:outer}). All remaining parameters, including the RV jitter, offset, and semi-amplitudes, were assigned broad, uniform priors.

We performed the RV-only fit using the \texttt{PyMC} framework in conjunction with the RV models from the \texttt{exoplanet} package. We used the same sampling strategy as in \S \ref{sec: transit}, employing four chains, each with 2000 tuning steps and 2000 draw steps. To compare the 18 models, we compute the Bayesian Information Criterion (BIC) for each model as
\begin{equation}
    \rm{ BIC} = k\ln(n) - 2\ln(\mathcal{L})
\end{equation}
where $\rm n$ is the number of data points, $\rm k$ is the number of free parameters, and $\mathcal{L}$ is the maximum likelihood. A model is favoured if it has a lower $\rm BIC$.

\begin{figure}[htbp]
    \centering
    \includegraphics[width=0.97\linewidth]{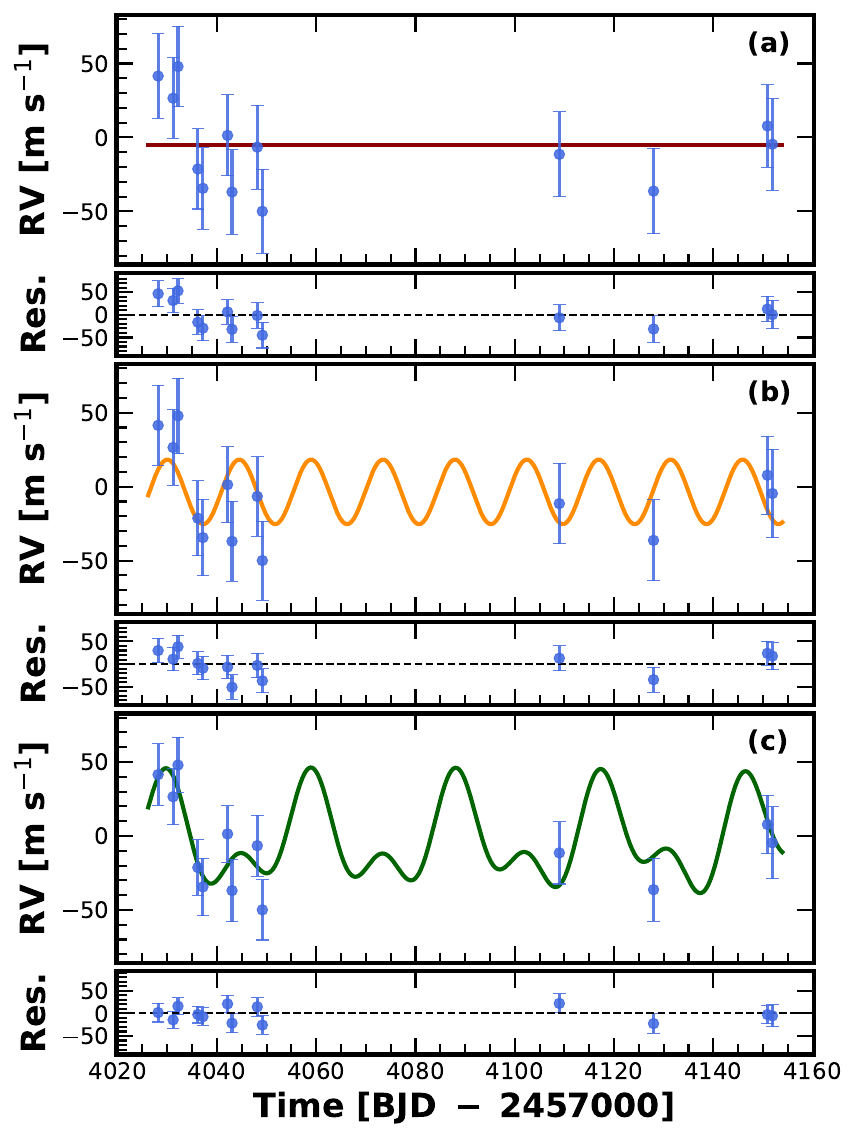}
    \caption{Radial velocity (RV) models and residuals fit to our MAROON-X observations. Here, we explore models consisting of (a) zero, (b) one, and (c) two planets to assess which configuration best explains the observed RV signal. The zero planet model consists exclusively of an RV offset and jitter while the 1- and 2-planet models contain Keplerian signals. The jitter term is added in quadrature to the observed MAROON-X uncertainties, resulting in varying error bars among the three panels. For the 2-planet model, we show the exterior perturber solution where the companion lies wide of the 2:1 MMR with TOI-3850.01. Both the residuals and the Bayesian Information Criterion (BIC) are minimized for the 2-planet model, suggesting an architecture where the perturbing planet lies exterior to TOI-3850.01 and near the 2:1 MMR.}
    \label{fig:rv_only}
\end{figure}

The result of each fit is shown in Table~\ref{tab:rv_bic_grid} while the median RV models for the 0-, 1- and 2-planet (assuming an exterior perturber near the 2:1 MMR) configurations are shown in Figure~\ref{fig:rv_only}. We find that the 2-planet model with the perturbing planet exterior to TOI-3850.01 and wide of the 2:1 MMR has the lowest BIC among the models tested. This architecture is consistent with the lack of observed transits for the companion, since an interior planet would be expected to transit in a nearly coplanar system. For the subsequent analysis, we assume the companion planet lies exterior to TOI-3850.01 and the two candidates are located near the 2:1 MMR.

\subsection{Joint RV + TTV Fit}
\label{sec:joint}

After establishing that the RV data support a two-planet solution near the 2:1 MMR, we perform a joint fit to the RV and TTV data to determine the final orbital and planetary parameters of the system. Modeling the RVs and TTVs simultaneously provides a self-consistent determination of the system parameters.

We model the observed transit times from \S \ref{sec: transit} using the \texttt{TTVFast} Python package \citep{deck2014}. \texttt{TTVFast} is a symplectic N-body integrator specifically designed to find the mid-transit times in dynamically interacting multi-planet systems. Instead of using \texttt{exoplanet} for RV modeling, we use the \texttt{RadVel} package \citep{fulton2018}. This approach is necessary since \texttt{TTVFast} is not readily compatible with the sampling methods used by \texttt{PyMC}. We parametrize the model in terms of the orbital periods, transit epochs, RV semi-amplitudes, relative longitude of ascending node, inclination angles, and eccentricity vector components $(h,k) = (\sqrt{e}\cos \omega, \sqrt{e} \sin \omega)$. We also fit for an RV offset and jitter term.

 Although we do not have a constraint on the inclination of the companion, we assume the system is nearly coplanar, as is commonly observed in multi-planet systems \citep{lissauer2011}. Furthermore, we do not observe any transits of the perturbing planet, indicating that it is a non-transiting companion. At exact coplanarity with TOI-3850.01, the companion would have an impact parameter greater than unity and would therefore not transit. 
 
In our dynamical fits, we impose a Gaussian prior on the inclination of TOI-3850.01 from our transit-only fit and allow the inclination of the companion to vary within $70^{\circ}<i_2<90^{\circ}$. Due to the absence of any observed transits, we require that the impact parameter of the perturbing planet be greater than unity. We set the longitude of the ascending node of TOI-3850.01 to $\Omega_1 = 0^{\circ}$ while placing a Gaussian prior on the node of the companion centered around $\Omega_2 = 0^{\circ}$. To mitigate the known mass-eccentricity degeneracy in TTV fits \citep{hadden2016}, we impose a Gaussian prior on the eccentricity components $(h,k)$ centered about 0. All remaining parameters are assigned physically-motivated uniform priors (see Table~\ref{tab:priors} for summary of priors).

The TTV posterior for systems with non-transiting planets is often highly degenerate, as multiple dynamically distinct solutions can reproduce the same observed timing signal \citep{lammers2026, boue2012}. We therefore first perform an exploratory search of the joint RV+TTV posterior with \texttt{pyABC}, a likelihood-free sampler based on approximate Bayesian computation \citep{schalte2022}, to identify high-probability regions of parameter space. The resulting maximum a posteriori (MAP) solution is then used to initialize the walkers of the MCMC. This two-step process reduces the risk of initializing the chains in areas of low probability or missing the dominant mode.

We used an MCMC within the \texttt{emcee} package to sample the final posterior distribution. 60 walkers were used to explore the 15-dimensional parameter space. The sampler was run for a total of 100,000 steps per walker and the first 25,000 were discarded as burn-in. The post-burn-in chains were run for more than 50 times the estimated autocorrelation time to ensure convergence. These remaining samples were then used to compute the best-fit values, for which we report the median values and the $1\sigma$ uncertainties corresponding to the 16th and 84th percentiles.

\begin{figure*}[htbp]
    \centering
    \includegraphics[width=0.99\linewidth]{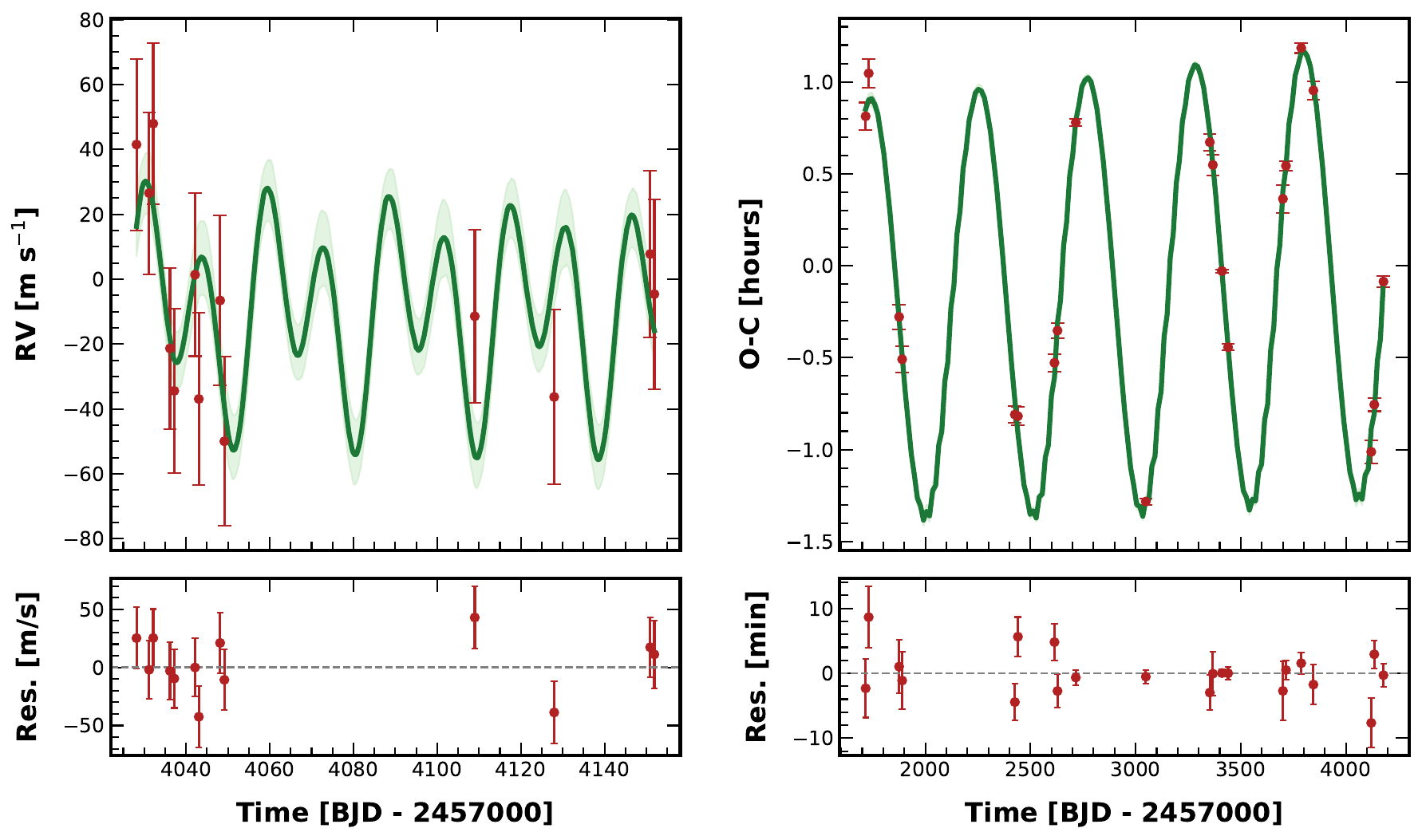}
    \caption{Best-fit joint radial velocity (\emph{left}) and transit timing variation (\emph{right}) model for the TOI-3850 system, suggesting the presence of two planets: TOI-3850 b and TOI-3850 c. TOI-3850 b is a Saturn-mass planet with $M_b \approx 112~M_{\oplus}$ at an orbital period of $P_b \approx 14.5~\mathrm{days}$, while TOI-3850 c is a non-transiting, Saturn-mass companion with $M_c \approx 90~M_{\oplus}$, located just wide of the 2:1 mean motion resonance (MMR) at $P_c \approx 29.8~\mathrm{days}$. The residuals for each fit are depicted in the bottom panels. For parts of the fit, the shaded regions depict the $1\sigma$ confidence interval.}
    \label{fig:rv+ttv}
\end{figure*}

The best-fit RV and TTV models are shown in Figure~\ref{fig:rv+ttv} while the inferred planetary parameters are summarized in Table~\ref{tab:planet_params}. Our joint analysis of the RVs and TTVs confirms the presence of a pair of planets located just wide of the 2:1 MMR. We henceforth designate these planets as TOI-3850 b and TOI-3850 c. TOI-3850 b $(P_b=14.484~\mathrm{days},~ M_b =112\pm20~M_{\oplus},~e_b = 0.018\pm0.008, R_b = 12.07\pm0.09~R_{\oplus}, ~T_{\rm{eq}}=841\pm10~\rm{K})$ is a transiting warm Jupiter and TOI-3850 c $(P_c=29.85\pm0.01~\mathrm{days},~ M_c =90\pm15~M_{\oplus},~e_c < 0.015, ~T_{\rm{eq}}=661\pm7~\rm{K})$ is a non-transiting, Saturn-mass companion located just wide of the 2:1 MMR. Because the joint RV+TTV analysis constrains the orbital inclination of TOI-3850 c, both inferred masses are the true dynamical masses rather than the usual RV minimum mass, $M_p \sin i$.

\begin{table*}[htbp]
\centering
\caption{System parameters for TOI-3850 b and c from the transit-only and joint RV+TTV fits. Reported values correspond to the median of the posterior distributions with 16th and 84th percentile uncertainties. All orbital parameters from the joint RV+TTV fit are osculating astrocentric elements evaluated at the reference epoch $t_{\rm ref} = \mathrm{1715}~{\rm TBJD}$.}

\label{tab:planet_params}
\vspace{0.5em}
\renewcommand{\arraystretch}{1.15}
\begin{tabular*}{0.95\textwidth}{@{\extracolsep{\fill}}llll@{}}
\toprule
Parameter & TOI-3850 b & TOI-3850 c & Description \\
\midrule

\multicolumn{4}{c}{\textit{Transit-Only Fit}} \\
\midrule
$P$ (days) & $14.48385 \pm 0.00001$ & -- & Orbital period \\
$T_0$ (TBJD)$^{a}$ & $1716.241 \pm 0.003$ & -- & Transit epoch \\
$R_p/R_\star$ & $0.107 \pm 0.001$ & -- & Radius ratio \\
$b$ & $0.680 \pm 0.006$ & -- & Impact parameter \\
$i$ (deg) & $88.44 \pm 0.01$ & -- & Inclination \\

\midrule
\multicolumn{4}{c}{\textit{Joint RV + TTV Fit}} \\
\midrule
$P$ (days) & $14.48402 \pm 0.00228$ & $29.84662 \pm 0.01026$ & Orbital period \\
$T_0$ (TBJD) & $1716.242 \pm 0.002$ & $4011.475 \pm 0.434$ & Transit epoch \\
$K$ ($\mathrm{m~s^{-1}}$) & $27.6\pm4.9$ & $17.5 \pm 2.9$ & RV semi-amplitude \\
$i$ (deg) & $88.44 \pm 0.02$ & $87.18 \pm 1.29$ & Inclination \\
$\Omega$ (deg) & -- & $3.82 \pm 2.87$ & Longitude of the ascending node \\

\midrule
\multicolumn{4}{c}{\textit{Derived Planetary Parameters}} \\
\midrule
$e$ & $0.018\pm0.008$ & $e_c < 0.015$ & Eccentricity \\
$\omega$ (deg) & $214\pm22$ & $228\pm92$ & Argument of pericenter \\

$R_p$ ($R_\oplus$) & $12.07 \pm 0.09$ & -- & Planet radius \\
$M_p$ ($M_\oplus$) & $112\pm20$ & $90\pm15$ & Planet mass \\
$T_{\rm{eq}}$ (K) & $841 \pm 10$ & $661\pm7$ & Equilibrium temperature \\ 
$a$ (au) & $0.12004 \pm 0.00072$ & $0.19435 \pm 0.00119$ & Semi-major axis \\
$\rho_p$ ($\mathrm{g~cm^{-3}}$) & $0.351\pm0.061$ & -- & Planet density \\
$\Delta i$ (deg) & -- & $4.58\pm 1.70$ & Mutual inclination \\  

\midrule
\multicolumn{4}{c}{\textit{RV Parameters}} \\
\midrule
$\mu_{\rm MAROON-X}$ ($\mathrm{m~s^{-1}}$) & \multicolumn{2}{c}{$-9.9\pm7.9$} & MAROON-X RV offset \\
$\sigma_{\rm MAROON-X}$ ($\mathrm{m~s^{-1}}$) & \multicolumn{2}{c}{$23.6 \pm 6.6$} & MAROON-X RV jitter \\

\bottomrule
\end{tabular*}

\vspace{0.5em}
\begin{minipage}{0.95\textwidth}
\footnotesize
$^{a}$ TBJD is BJD $-2457000$.
\end{minipage}

\end{table*}

\section{Discussion} \label{sec:discussion}

\subsection{Architecture of the TOI-3850 System}
\label{sec:orb_arch}

\subsubsection{Formation History of TOI-3850}

\begin{figure}[htbp]
    \centering
    \includegraphics[width=0.95\linewidth]{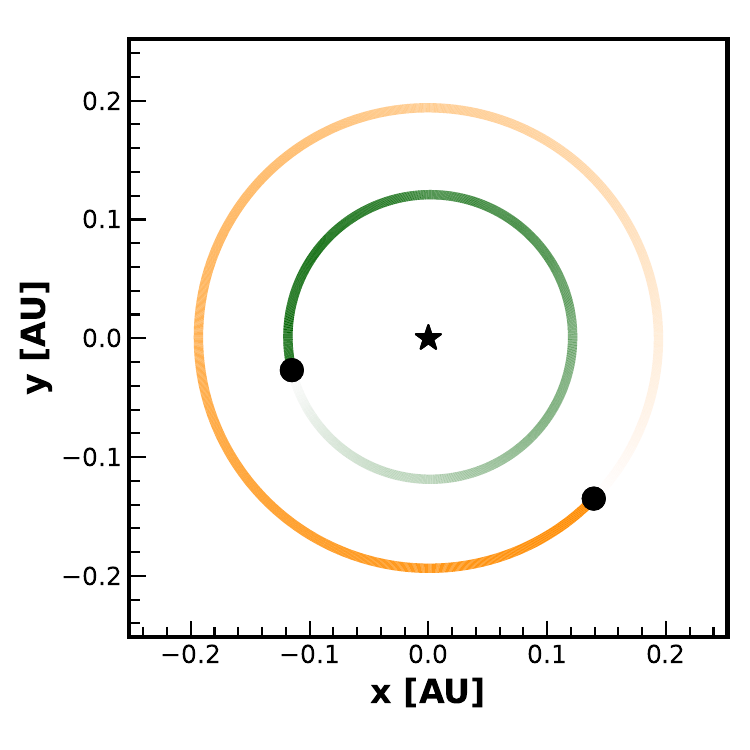}
    \caption{Depiction of the orbits of TOI-3850 b and TOI-3850 c based on the median values in the joint RV + TTV fit.}
    \label{fig:orbit}
\end{figure}

In \S \ref{sec:joint}, we detect the presence of two planets on near-circular orbits around TOI-3850: TOI-3850 b $(P_b=14.484~\mathrm{days},~ M_b =112\pm20~M_{\oplus})$, a transiting warm Jupiter, and TOI-3850 c $(P_c=29.85\pm0.01~\mathrm{days},~ M_c =90\pm15~M_{\oplus})$, a Saturn-mass, non-transiting companion. The orbits of the two planets are shown in Figure~\ref{fig:orbit}. The two planets are found on compact orbits within $0.2~\rm AU$ and have a period ratio of $P_c/P_b \approx2.06$. It is unsurprising that the two planets are found \textit{outside} resonance, as pairs of planets are frequently observed just wide of exact MMR commensurabilities \citep{lithwick2012}. This offset is commonly attributed to dissipative forces, such as tides, which damp eccentricities and drive planets away from exact resonance \citep{batygin2012}. To place the TOI-3850 system in broader context, we find that only eleven confirmed systems have multiple Saturn-mass planets $(M_p>0.25M_{\rm J})$ with orbital separations less than 0.2 AU, despite the relative ease of detecting and confirming such pairs of planets.

The compact, near-resonant architecture of the TOI-3850 system provides important insights into its formation history. It is unlikely that the two planets formed in situ at semi-major axes $< 0.2~\rm AU$; it is expected that the local surface density in protoplanetary disks is too low to form multiple massive cores required for giant planet formation \citep{ford2014}.  Furthermore, models of in situ formation for short-period gas giants generally predict systems with low-mass companions, rather than multiple giants \citep{batygin2016}. On the other hand, it is also unlikely the two planets underwent high-eccentricity migration, as this pathway typically results in the ejection of nearby companion planets, leaving behind either an isolated gas giant or companions on more distant orbits \citep{dawson2018}. This scenario has been invoked to explain the observed lack of nearby companions found around hot Jupiters \citep{mustill2015}. 

Due to the system's proximity to the 2:1 MMR, it is most naturally explained by disk-driven migration \citep{lee2002}. In this scenario, TOI-3850 b and TOI-3850 c likely formed near or beyond the snowline and subsequently migrated inwards through gravitational interactions with the protoplanetary disk. As long as the mass of the outer companion is comparable to the mass of the inner companion, two planets can undergo convergent type I disk migration \citep{lin2025}. Eventually, the period ratio approaches a first-order MMR commensurability and the planets become trapped in resonance through repeated gravitational interactions.

\subsubsection{Comparison to Other Warm-Jupiter Systems}

Only a few compact systems with multiple giant planets similar to TOI-3850 have been found. For example, the TOI-1232 system hosts a pair of nearly coplanar giant planets near the 2:1 MMR \citep{mihaylov2026}. Similar to TOI-3850, TOI-1232 hosts a transiting warm Jupiter at a period of $P_b\approx14.3~\rm days$ and a non-transiting companion at $P_c\approx30.4~\rm days$. However, these two planets have a slightly higher period ratio of $P_c/P_b \approx 2.13$, placing them further from resonance. In addition, the planets in TOI-3850 are more closely spaced and exhibit more comparable masses. Other examples of multi-planet warm Jupiter systems include Kepler-108, HAT-P-17, and TOI-7510 \citep{mills2017, howard2012, almenara2025}, whose architectures suggest relatively quiescent dynamical histories, such as disk-driven migration.

Warm Jupiters may represent a heterogeneous population with multiple formation and migration pathways. Several warm Jupiters have been found in single planet systems (e.g., WASP-130 b and TOI-5027 b; \citealt{hellier2017,pinto2025}) with many displaying elevated eccentricities (e.g., TOI-6019 b and TIC-241249530 b ; \citealt{thomas2026, gupta2024}). These apparently isolated and eccentric systems are often interpreted as products of high-eccentricity migration, in contrast to the more dynamically quiescent histories, such as disk-driven migration, inferred for many multi-planet warm-Jupiter systems \citep{huang2016, petrovich2016}. In this picture, it is possible that a subpopulation of lone warm Jupiters are progenitors of hot Jupiters while others may have stalled at warm Jupiter distances. However, low-eccentricity lone warm Jupiters may form through either quiescent or high-eccentricity migration pathways. 

\subsection{Dynamics of the TOI-3850 System}

\subsubsection{Dynamical Stability}
\label{subsec: stability}

We first tested the dynamical stability of the TOI-3850 system using the \texttt{SPOCK} stability classifier \citep{tamayo2020}. \texttt{SPOCK} is a machine learning model that computes the probability that a given orbital configuration will remain stable over the first 1 billion orbits. Although the model was trained exclusively on near-resonant 3-planet systems, in the two-planet regime, \texttt{SPOCK} effectively behaves like a binary classifier. This is consistent with the sharp stability boundary produced by two-body MMR overlap \citep{tamayo2021}. To assess stability, we randomly drew 100,000 samples from the posterior distribution of the joint RV+TTV fit. For each sample, we recorded the stability probability and found that 99.33\% of the draws remain stable for the first $10^9$ orbits. Unstable configurations are typically associated with mass ratios deviating far from unity $(M_c/M_b\gg1~\mathrm{or}~M_c/M_b\ll1)$ or elevated eccentricities $(e_b \gtrsim 0.1)$ for TOI-3850 b.

To further validate the stability of the system, we performed direct $N$-body integrations using the \texttt{REBOUND} package \citep{rein2012}. We randomly drew 1000 samples from the posterior distribution of the joint RV+TTV fit and ran the simulations for 250,000 years. Each integration was performed with the \texttt{WHFast} symplectic integrator with a time step of $\Delta t = P_b/5$. We classify configurations as unstable if they exhibit orbit crossing or if either planet is ejected beyond $0.6~\rm AU$. In addition, we recorded the mutual Hill radius
\begin{equation}
    R_{H} =\left(\frac{M_b+M_c}{3M_{\star}} \right)^{1/3} \left(\frac{a_b+a_c}{2} \right)
\end{equation}
and the orbital separation in units of their Hill radius,
\begin{equation}
    \Delta = (a_c-a_b)/R_H
\end{equation}
where $a_b,~a_c$ are the semi-major axes of TOI-3850 b and TOI-3850 c, respectively \citep{gratia2021}. Following \citet{gladman1993}, we require stable configurations satisfy $\Delta > 2\sqrt{3}$.

Out of the 1000 posterior realizations, $99.3\%$ did not trigger any instability criteria and thus, remained dynamically stable over 250,000 years. The seven unstable samples resulted in the ejection of TOI-3850 c and had elevated eccentricities for both planets, suggesting instability is strictly confined to the high-eccentricity tail of the posterior. These seven samples also had minimum separations of $\Delta \approx 6$, indicating close encounters can lead to ejection of the outer planet.

Together, the \texttt{SPOCK} classification and direct $N$-body integrations imply the joint RV+TTV solution for TOI-3850 is stable across the majority of the posterior.

\subsubsection{Orbital Evolution}

While many compact multi-planet systems are found near first-order MMR commensurabilities, proximity to a period ratio does not imply resonant locking \citep{macdonald2023, goldreich2014}. To examine whether the planets are typically trapped in an exact 2:1 MMR, we track the first-order resonant angles 
\begin{equation}
    \theta_1 = 2\lambda_c - \lambda_b - \varpi_b
\end{equation}
\begin{equation}
    \theta_2 = 2\lambda_c - \lambda_b - \varpi_c 
\end{equation}
where $\lambda = M + \omega + \Omega $ is the mean longitude and $ \varpi = \omega + \Omega$ is the longitude of periastron. Following \citet{mustill2011} and \citet{lee2002}, two planets are locked in exact resonance only if one of the two resonant angles librates around a fixed value. If both angles circulate between $-180^{\circ}$ and $180^{\circ}$, then the planets are near the commensurability but not trapped in resonance.

Using the same posterior samples and procedure as \S \ref{subsec: stability}, we integrated each system for 1000 years, recorded the first-order resonant angles and classified each system as circulating or librating. From these integrations, we also estimated the mean orbital periods by averaging the osculating mean motions, finding $\bar{P}_b = 14.48333 \pm 0.00243~\rm days$ and $\bar{P}_c = 29.89736 \pm  0.01907 ~ \rm days$. Out of the 1000 samples, we determine that 8.8\% display $\theta_1$ libration about $0^{\circ}$, suggesting that a small but non-negligible fraction of the posterior solutions are consistent with resonant behaviour in the exact 2:1 MMR. 

\begin{figure}[htbp]
    \centering
    \includegraphics[width=0.95\linewidth]{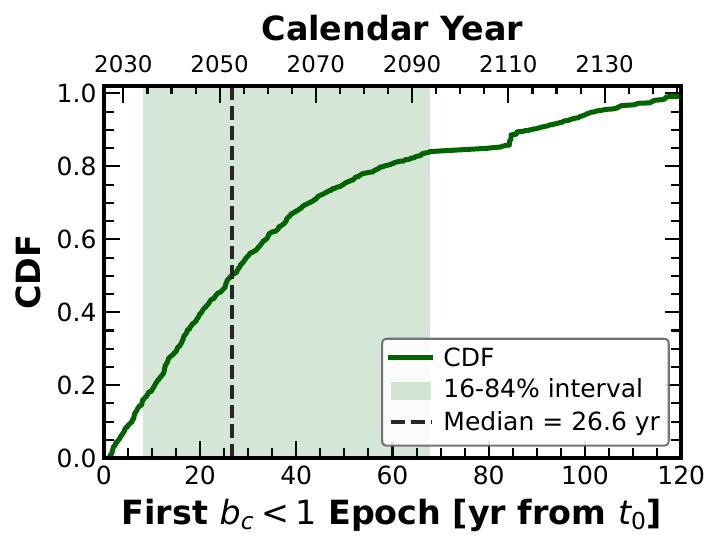}
    \caption{Cumulative distribution function (CDF) of the first time TOI-3850 c reaches a transit-permitting geometry, defined as $b_c < 1$, measured from $t_0 = 4010$ TBJD (2025 November 30). The dashed line marks the median (26.6 yr) and the shaded region shows the 16th--84th percentile interval.}
    \label{fig:transit_cdf}
\end{figure}

For each timestep, we also recorded the impact parameter of both TOI-3850 b and TOI-3850 c. We find that 50\% of samples reach a transiting geometry for TOI-3850 c within the next $\sim27$ years, defined conservatively by $b_c<1$ during the simulations (see Figure~\ref{fig:transit_cdf}). Although TOI-3850 c is currently inferred dynamically through RVs and TTVs, these simulations indicate that roughly half of the posterior solutions predict transit events in the near future.

\begin{figure*}[htbp]
    \centering
    \includegraphics[width=0.85\linewidth]{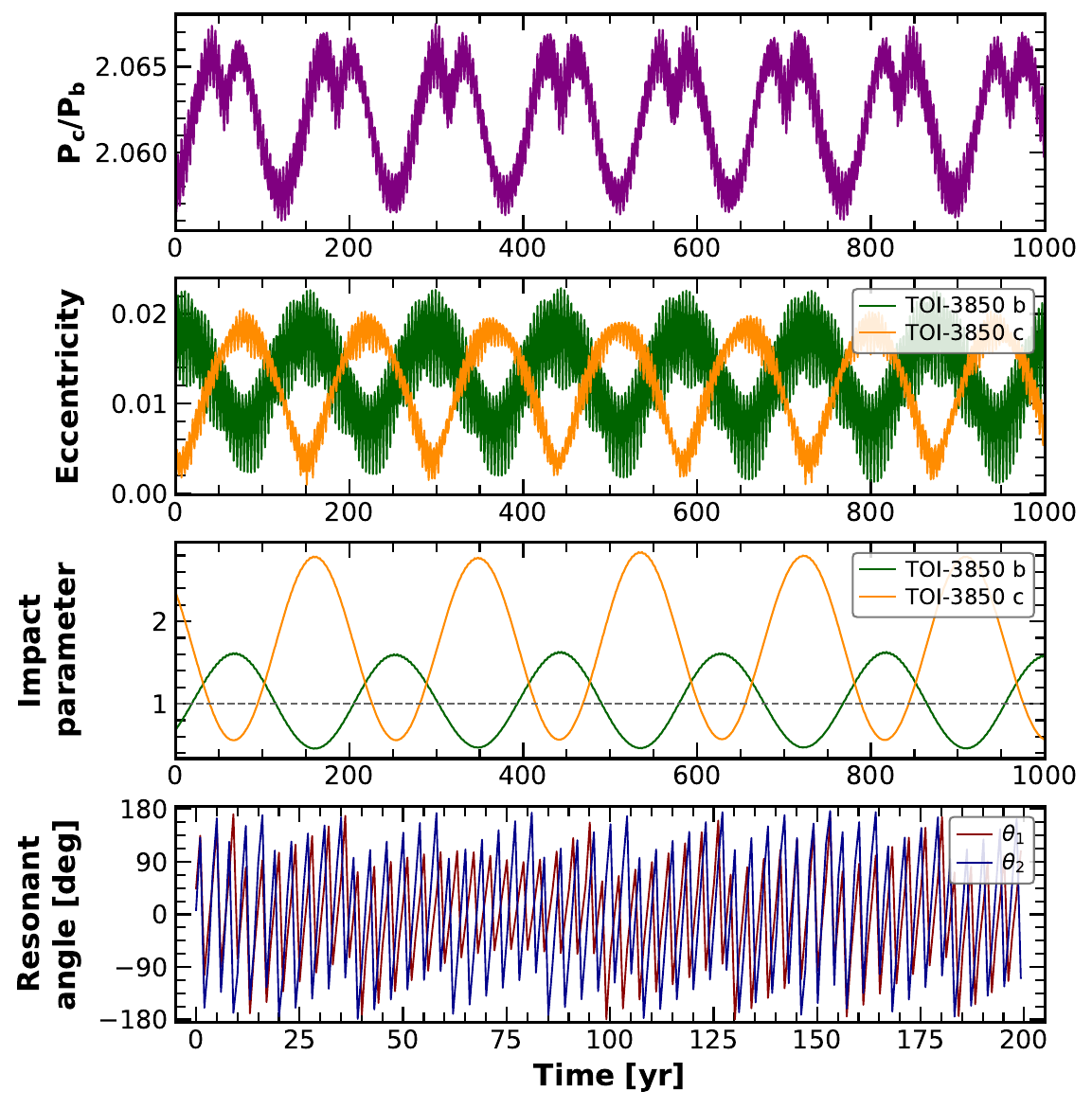}
    \caption{Time evolution of the best-fit dynamical configuration of the TOI-3850 system from a $1~\rm Myr$ integration with \texttt{REBOUND}. The osculating period ratio, eccentricities, and impact parameters are shown for the first 1000 years while the resonant angles are shown for the first 200 years. We do not observe any unbounded eccentricity growth that could lead to chaos or orbit crossings and the period ratio remains bounded well above the exact 2:1 commensurability. The evolution of the impact parameters suggest that TOI-3850 c could reach a transiting configuration in the near future, while TOI-3850 b may become non-transiting. Both resonant angles circulate between $-180^{\circ}$ and $180^{\circ}$, indicating that the best-fit configuration is near, but not necessarily locked in, the exact 2:1 MMR.}
    \label{fig:orb_elm}
\end{figure*}

We further integrated the 88 samples displaying libration, as well as the median system parameters, over $1~\rm Myr$ in \texttt{REBOUND} to investigate long-term behaviour. Using the same three criteria as \S \ref{subsec: stability}, we find that the median parameters and 96.6\% of the initially librating samples remain stable throughout the integration, with the orbital elements displaying bounded, quasi-periodic variations (see Figure~\ref{fig:orb_elm}). However, the resonant behaviour is not always persistent as 77\% of the initially librating samples alternate between libration and circulation at least once in the 1 Myr simulation. It is therefore plausible that a subset of the initially circulating samples may transition between librating and circulating on longer timescales.

While chaos is often associated with instability, the orbital elements do not display any chaotic behaviour over the probed timescale (see Figure~\ref{fig:orb_elm}). Specifically, orbits with unbounded eccentricity growth can lead to orbit crossings and the eventual ejection of one or more planets \citep{marzari2014}. Here, we see both planets have near-zero, bounded eccentricities. We also observe that the orbital separation remains above the critical limit of $\Delta \approx 3.46$ in the majority of simulations.

In Figure~\ref{fig:orb_elm}, we observe that both $\theta_1$ and $\theta_2$ circulate between $-180^{\circ}$ and $180^{\circ}$ for the median system parameters, rather than librating about a fixed value. Similarly, the osculating period ratio remains bounded above exact 2:1 commensurability. Combined with the classification of the posterior samples above, these results suggest that TOI-3850 b and TOI-3850 c are most likely near, but not exactly trapped in, the 2:1 MMR.

\subsection{Potential for Follow-up}
\label{sec:followup}

\subsubsection{Atmospheric Characterization}
\begin{figure}[htbp]
    \centering
    \includegraphics[width=1\linewidth]{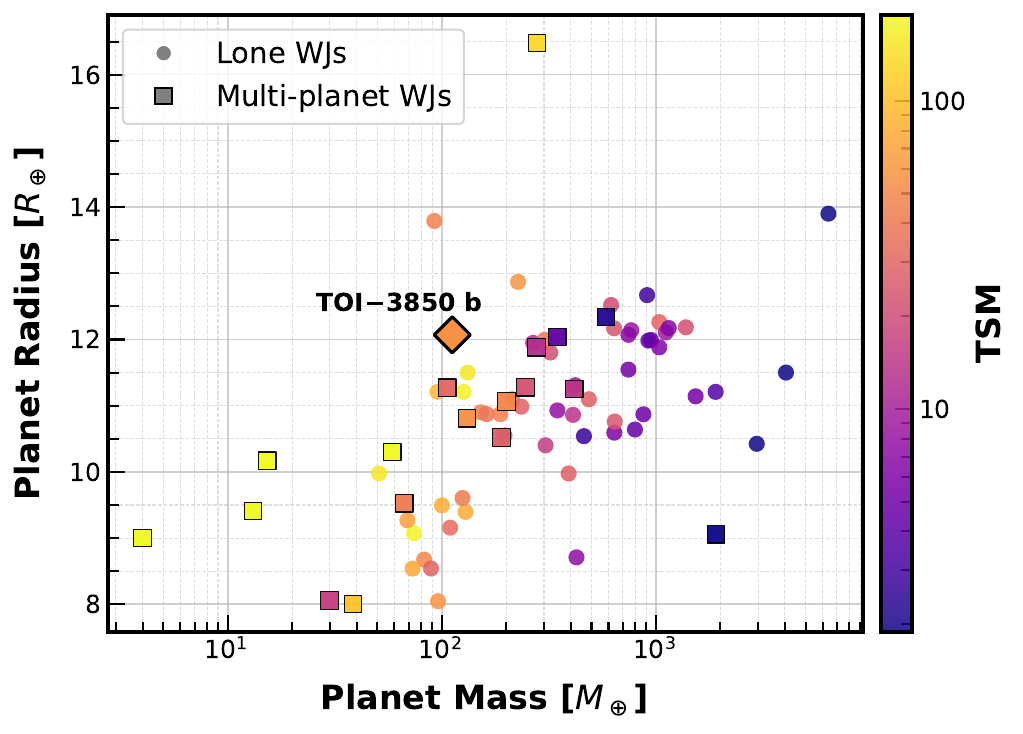}
    \caption{Radius-mass diagram of all confirmed warm Jupiters with the colour scale set by their transmission spectroscopy metric (TSM). All lone warm Jupiters are shown as circles while warm Jupiters in multi-planet systems are depicted as squares. TOI-3850 is shown as the only diamond. Out of the 40 confirmed multi-planet warm Jupiters, TOI-3850 b has the seventh highest TSM, indicating it is a strong target for follow-up atmospheric characterization.}
    \label{fig:tsm}
\end{figure}

To place TOI-3850 b in the context of other warm Jupiter atmospheres, we computed its transmission spectroscopy metric (TSM, \citealt{kempton2018}). The TSM is the expected SNR of a 10 hr transmission spectroscopy observation with JWST/NIRISS. We find that TOI-3850 b has a TSM of 58, which is higher than $87\%$ of all confirmed warm Jupiters. When compared specifically to warm Jupiters in multi-planet systems, we find TOI-3850 b ranks seventh highest in TSM. This places TOI-3850 b among the most favourable targets within its population, making it a strong candidate for atmospheric characterization. In Figure~\ref{fig:tsm}, we show all confirmed warm Jupiters on a radius-mass diagram, with the colour scale denoting TSM.

TOI-3850 b is a compelling target for future atmospheric characterization with the Ariel mission. With a launch scheduled for 2031, the Ariel mission will survey the atmospheres of roughly 1000 exoplanets \citep{tinetti2022}. Because some warm Jupiters may share formation pathways with hot Jupiters, the population may consist of distinct subclasses: those that form via similar mechanisms to hot Jupiters and those that form through alternative pathways \citep{dawson2018}. By enabling comparative atmospheric studies, Ariel will be able to systematically test whether lone warm Jupiters and those in multi-planet systems share similar atmospheric properties. These observations will consequently provide constraints on their formation and evolution histories. Given that it has been recommended Ariel targets $\sim20$ warm Jupiters \citep{daoust2025}, we expect that TOI-3850 b will be selected for atmospheric follow-up.

\subsubsection{Gaia Astrometry}
Based on 66 months of astrometric observations, Gaia DR4, currently expected in December 2026, will enable the detection of massive, long-period exoplanets typically inaccessible to transit and RV surveys \citep{vallenari2023, perryman2014}. To evaluate the strength of the astrometric signals of TOI-3850 b and c, we calculate the angular semi-major axis, or \textit{astrometric signature}, as
\begin{equation}
\alpha =
\left(\frac{M_p}{M_\star}\right)
\left(\frac{a_p}{1~\mathrm{AU}}\right)
\left(\frac{d}{1~\mathrm{pc}}\right)^{-1}
\mathrm{arcsec}
 \label{eq:astrometric_sig}
\end{equation}
to obtain $\alpha_b \approx 0.08~\rm{\mu as}$ and $\alpha_c \approx0.1~\rm{\mu as}$ \citep{perryman2014}. Following \citet{casertano2008} and assuming a single-measurement error of $\sigma_{\rm{single}} = 50~\rm{\mu as}$, we classify planets with astrometric signatures $\alpha \gtrsim 150~\rm{\mu as}$ as recoverable with Gaia. Both TOI-3850 b and c remain well below this detection limit and thus will not be recoverable with Gaia astrometry. 

By rearranging Equation~(\ref{eq:astrometric_sig}) and invoking Kepler's Third Law, one can solve for the minimum orbital period observable with Gaia
\begin{equation}
    P_{\rm{min}} = 2\pi \sqrt{\left (\frac{1}{G(M_p+M_{\star})} \right) \left( \frac{d M_{\star}\alpha_{\rm{min}}}{M_p} \right)^3}
\end{equation}
where $\alpha_{\rm{min}} = 150~\rm{\mu as}$ is our assumed detection sensitivity. Since Gaia DR4 will span 66 months of observations, the maximum well-sampled orbital period is $P_{\rm{max}} \approx 5.5~\rm{yr}$ \citep{casertano2008}, which corresponds to a low-mass brown dwarf with $M_{BD} \approx 26~{M_J}$. On the other hand, for a brown dwarf near the hydrogen-burning limit with $M_{BD} = 80~M_J$, we find $P_{\rm{min}} \approx 1~\rm{yr}$. Therefore, Gaia DR4 will be sensitive to brown dwarfs with approximate orbital periods of $1.0~\mathrm{yr}~ \lesssim P \lesssim 5.5~\mathrm{yr}$ around TOI-3850. No planetary-mass companion reaches the adopted detection threshold within this period range.

\section{Summary} \label{sec:conclusion}

Overall, in this work, we study a multi-planet warm Jupiter system around TOI-3850 using transit photometry, stellar spectroscopy, and transit timing variations. After characterizing the star through periodogram analysis, isochrone fitting, and reconnaissance spectroscopy, we find TOI-3850 is consistent with a young $\left (\tau_{\star} = 0.89^{+1.01}_{-0.63}~\rm Gyr \right)$, relatively fast-rotating $(P_{\rm rot} \approx 6~\rm days)$, moderately active, and near-solar metallicity $\left ([\rm Fe/H] = 0.10\pm0.05 \right)$ G0 dwarf star. While TOI-3850 was first observed by TESS, follow-up ground-based photometric observations enabled precision timing measurements, revealing the presence of a non-transiting outer companion planet.

By jointly fitting the observed TTVs and RVs to a single, self-consistent dynamical model, we detect the presence of two giant planets located on near-circular, coplanar orbits just wide of the 2:1 MMR $(P_c/P_b \approx2.06)$. TOI-3850 b $(P_b=14.484~\mathrm{days},~ M_b =112\pm20~M_{\oplus},~e_b = 0.018\pm0.008, R_b = 12.07\pm0.09~R_{\oplus}, ~T_{\rm{eq}}=841\pm10~\rm{K})$ is a transiting Saturn-mass warm Jupiter while TOI-3850 c $(P_c=29.85\pm0.01~\mathrm{days},~ M_c =90\pm15~M_{\oplus},~e_c < 0.015, ~T_{\rm{eq}}=661\pm7~\rm{K})$  is a non-transiting, Saturn-mass companion.

Our dynamical analysis shows that the system is near, but probably not locked in the exact 2:1 MMR. In particular, we find that TOI-3850 b and TOI-3850 c are likely located exterior to resonance rather than undergoing resonant libration. The compact architecture of the TOI-3850 system thus provides an important laboratory for testing how  giant planets interact, migrate, and escape exact MMR commensurability. 

Out of the more than $6000$ confirmed exoplanets, only eleven systems have multiple giant planets $(M_p>0.25~M_{\rm J})$ with orbital separations less than 0.2 AU, demonstrating the unusual nature of compact systems such as TOI-3850. TOI-3850 therefore joins a small but valuable population of compact systems that can be used to constrain the formation and evolution of giant planets with massive, nearby companions.

\section*{Acknowledgments}

This paper includes data collected by the TESS mission that are publicly available from the Mikulski Archive for Space Telescopes (MAST). Funding for the TESS mission is provided by NASA's Science Mission Directorate. This research has made use of the Exoplanet Follow-up Observation Program (ExoFOP; DOI: 10.26134/ExoFOP5) website, which is operated by the California Institute of Technology, under contract with the National Aeronautics and Space Administration under the Exoplanet Exploration Program. We acknowledge the use of public TESS data from pipelines at the TESS Science Office and at the TESS Science Processing Operations Center. Resources supporting this work were provided by the NASA High-End Computing (HEC) Program through the NASA Advanced Supercomputing (NAS) Division at Ames Research Center for the production of the SPOC data products.

This work makes use of observations from the LCOGT network. Part of the LCOGT telescope time was granted by NOIRLab through the Mid-Scale Innovations Program (MSIP). MSIP is funded by NSF. This paper is based on observations made with the Las Cumbres Observatory’s education network telescopes that were upgraded through generous support from the Gordon and Betty Moore Foundation. This paper is based on observations made with the MuSCAT3 instrument, developed by the Astrobiology Center and under financial supports by JSPS KAKENHI (JP18H05439) and JST PRESTO (JPMJPR1775), at Faulkes Telescope North on Maui, HI, operated by the Las Cumbres Observatory. 

Based on observations obtained at the international Gemini Observatory, a program of NSF NOIRLab, which is managed by the Association of Universities for Research in Astronomy (AURA) under a cooperative agreement with the U.S. National Science Foundation on behalf of the Gemini Observatory partnership: the U.S. National Science Foundation (United States), National Research Council (Canada), Agencia Nacional de Investigaci\'{o}n y Desarrollo (Chile), Ministerio de Ciencia, Tecnolog\'{i}a e Innovaci\'{o}n (Argentina), Minist\'{e}rio da Ci\^{e}ncia, Tecnologia, Inova\c{c}\~{o}es e Comunica\c{c}\~{o}es (Brazil), and Korea Astronomy and Space Science Institute (Republic of Korea). The Gemini observations were obtained under program IDs GN-2026A-FT-203 and GN-2025B-FT-109. We thank the MAROON-X team of the University of Chicago for providing radial velocity data. MK acknowledges the support of the Natural Sciences and Engineering Research Council of Canada (NSERC), RGPIN-2024-06452. Cette recherche a été financée par le Conseil de recherches en sciences naturelles et en génie du Canada (CRSNG), RGPIN-2024-06452. N.B.C. acknowledges support from an NSERC Discovery Grant, a Tier 2 Canada Research Chair, and an Arthur B. McDonald Fellowship.

The work of I.A.S. was conducted under the state assignment of Lomonosov Moscow State University. We acknowledge financial support from the Agencia Estatal de Investigaci\'on of the Ministerio de Ciencia e Innovaci\'on MCIN/AEI/10.13039/501100011033 and the ERDF “A way of making Europe” through projects PID2021-125627OB-C32 and PID2024-158486OB-C32. This work is supported by the ERC Grant (ERC Advanced Grant SPEAR, GA 101200674). Funded by the European Union. Views and opinions expressed are however those of the author(s) only and do not necessarily reflect those of the European Union or the European Research Council Executive Agency (ERCEA). Neither the European Union nor the granting authority can be held responsible for them. Funding for KB was provided by the European Union (ERC AdG SUBSTELLAR, GA 101054354). ChatGPT (OpenAI), using the GPT-5.6 model, was used to assist with proofreading and troubleshooting code. All suggested revisions were reviewed and verified by the authors.

\software{
\texttt{AstroImageJ} \citep{collins2017},
\texttt{Astropy} \citep{astropy2022},
\texttt{batman} \citep{kreidberg2015},
\texttt{ChatGPT} \citep{openai2026},
\texttt{emcee} \citep{foreman2013},
\texttt{exoplanet} \citep{foreman2021},
\texttt{isochrones} \citep{morton2015},
\texttt{lightkurve} \citep{lightkurve2018},
\texttt{LMFIT} \citep{newville2025},
\texttt{matplotlib} \citep{Hunter2007},
\texttt{NumPy} \citep{Harris2020},
\texttt{pandas} \citep{pandas2026},
\texttt{pyABC} \citep{schalte2022},
\texttt{PyMC} \citep{abrilpla2023},
\texttt{RadVel} \citep{fulton2018},
\texttt{REBOUND} \citep{rein2012},
\texttt{SciPy} \citep{Virtanen2020},
\texttt{SERVAL} \citep{zechmeister2018},
\texttt{SPOCK} \citep{tamayo2020},
\texttt{TTVFast} \citep{deck2014},
\texttt{Wotan} \citep{hippke2019}
}

\appendix
\setcounter{table}{0}
\renewcommand{\thetable}{A\arabic{table}}
\setcounter{figure}{0}
\renewcommand{\thefigure}{A\arabic{figure}}

\begin{figure*}[htbp]
    \centering
    \includegraphics[width=0.95\linewidth]{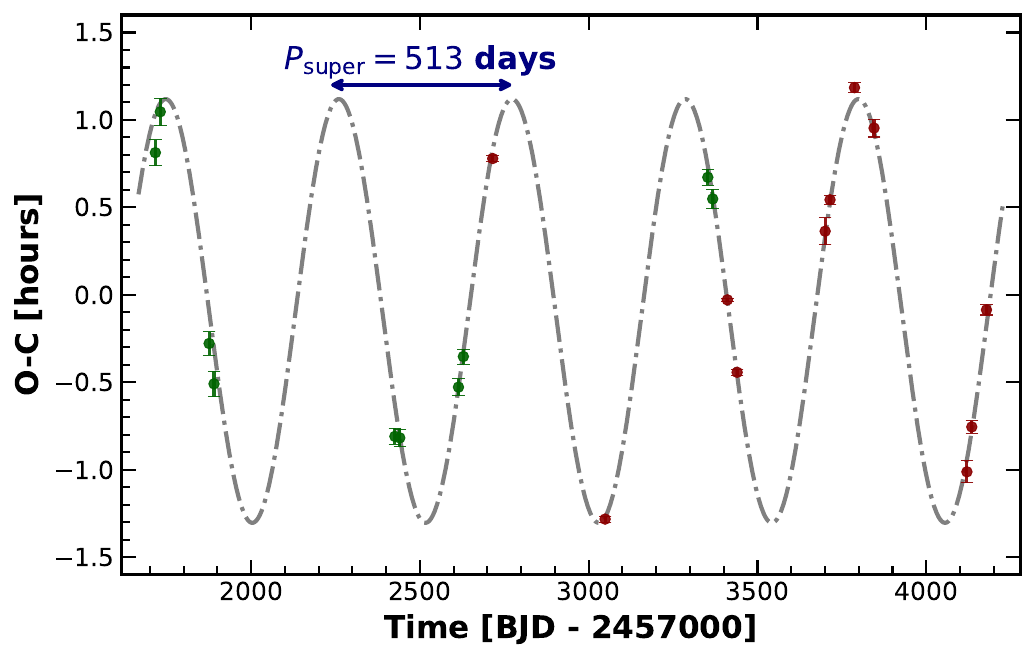}
    \caption{Transit timing variations (TTVs) of TOI-3850.01 from TESS (green) and ground-based telescopes (red), along with a sine wave model (dotted grey line). The sine model suggests the TTVs occur with a 513-day super period and an amplitude of 1.21 hours. The uncertainties on some of the ground-based TTVs are smaller than the data points shown. O–C (Observed minus Calculated) values show the difference between the observed mid-transit times and those predicted by a constant period model.}
    \label{fig:o-c}
\end{figure*}

In Table~\ref{tab:rv_measurements}, we record all MAROON-X RV observations. Figure~\ref{fig:o-c} shows the calculated TTVs from the transit-only fit in \S \ref{sec: transit} and Table~\ref{tab:transit_times} lists the derived mid-transit times of TOI-3850 b. Table~\ref{tab:rv_bic_grid} shows the results of the RV-only model comparison in \S \ref{sec:rv_only} while Table~\ref{tab:priors} states the priors used in the joint RV+TTV fit in \S \ref{sec:joint}.

\begin{table*}[htbp]
\centering
\caption{MAROON-X RV measurements of TOI-3850 from the red and blue arms. The reported uncertainties are the instrumental uncertainties and do not include the fitted RV jitter.}
\label{tab:rv_measurements}
\vspace{0.5em}
\renewcommand{\arraystretch}{1.15}
\begin{tabular*}{0.98\textwidth}{@{\extracolsep{\fill}}rrrrrc@{}}
\toprule
Epoch &
$\mathrm{RV}_{\mathrm{red}}$ &
$\sigma_{\mathrm{red}}$ &
$\mathrm{RV}_{\mathrm{blue}}$ &
$\sigma_{\mathrm{blue}}$ &
$t_{\mathrm{exp}}$ \\
$(\mathrm{BJD})$ &
($\mathrm{m~s^{-1}}$) &
($\mathrm{m~s^{-1}}$) &
($\mathrm{m~s^{-1}}$) &
($\mathrm{m~s^{-1}}$) &
(s) \\
\midrule
2461028.16 &  41.48 & 11.50 &  23.57 & 5.19 & 1000  \\
2461031.17 &  26.51 & 7.47 &  18.05 & 3.23 & 1000  \\
2461032.16 &  47.93 & 7.38 &  33.52 & 3.17 & 1000 \\
2461036.16 &  -21.33 & 6.96 &  -16.14 & 2.99 & 1000 \\
2461037.18 &  -34.42 & 8.78 &  -24.02 & 3.92 & 1000 \\
2461042.15 &  1.37 & 8.00 &  -22.73 & 3.54 & 1000 \\
2461043.05 &  -36.91 & 11.86 &  -6.81 & 5.25 & 1000 \\
2461048.13 &  -6.56 & 11.05 &  -21.62 & 4.85 & 1000 \\
2461049.15 &  -49.96 & 10.75 &  -32.78 & 4.73 & 1000 \\
2461103.95 &  109.04 & 15.18 &  128.93 & 6.59 & 1000 \\
2461108.98 &  -11.43 & 11.93 &  7.23 & 5.05 & 1000 \\
2461127.96 &  -36.29 & 12.41 &  -23.27 & 5.34 & 1000 \\
2461150.91 &  7.76 & 9.75 &  -20.69 & 4.16 & 1000 \\
2461151.90 &  -4.61 & 17.06 &  26.00 & 7.45 & 1000 \\
\bottomrule
\end{tabular*}

\end{table*}

\begin{table*}[htbp]
\centering
\caption{Measured mid-transit times of TOI-3850 b.}
\label{tab:transit_times}
\vspace{0.5em}
\renewcommand{\arraystretch}{1.15}

\hspace*{-\dimexpr 1in+\oddsidemargin\relax}%
\makebox[\paperwidth][c]{%
\begin{tabular}{ccc@{\hspace{3em}}ccc}
\toprule
Transit Number &
$T_{\mathrm{c}}$ &
$\sigma_{T_{\mathrm{c}}}$ &
Transit Number &
$T_{\mathrm{c}}$ &
$\sigma_{T_{\mathrm{c}}}$ \\
&
$(\mathrm{BJD}-2{,}457{,}000)$ &
(days) &
&
$(\mathrm{BJD}-2{,}457{,}000)$ &
(days) \\
\midrule
0   & 1716.24058257 & 0.00312817 &
114 & 3367.38832782 & 0.00232458 \\

1   & 1730.73416733 & 0.00327203 &
117 & 3410.81582714 & 0.00035870 \\

11  & 1875.51745359 & 0.00283732 &
119 & 3439.76625860 & 0.00068865 \\

12  & 1889.99169136 & 0.00304588 &
137 & 3700.50915752 & 0.00317450 \\

49  & 2425.88158057 & 0.00193672 &
138 & 3715.00049647 & 0.00103640 \\

50  & 2440.36506818 & 0.00210590 &
143 & 3787.44645617 & 0.00116454 \\

62  & 2614.18334627 & 0.00195105 &
147 & 3845.37222484 & 0.00213136 \\

63  & 2628.67448594 & 0.00178478 &
166 & 4120.48344788 & 0.00265541 \\

69  & 2715.62475578 & 0.00079243 &
167 & 4134.97799538 & 0.00150614 \\

92  & 3048.66736854 & 0.00073177 &
170 & 4178.45742553 & 0.00122071 \\

113 & 3352.90962838 & 0.00186767 &
    &               &            \\
\bottomrule
\end{tabular}%
}

\end{table*}

\begin{table*}[htbp]
\centering
\caption{Model comparison results for the RV-only fits to the 0-, 1- and 2-planet configurations. Reported $\Delta$BIC values are reported relative to the exterior model located wide of the 2:1 MMR.}
\label{tab:rv_bic_grid}
\vspace{0.5em}
\renewcommand{\arraystretch}{1.15}
\begin{tabular*}{0.98\textwidth}{@{\extracolsep{\fill}}llllccc@{}}
\toprule
Model & Architecture & MMR & Branch & $P_2$ Prior (days) & BIC & $\Delta$BIC \\
\midrule
0-planet & -- & -- & -- & -- & 130.3 & 6.6 \\
1-planet & -- & -- & -- & -- & 129.2 & 5.5 \\

\midrule
2-planet & Exterior & 2:1 & Shortward $(+)$ & $\mathcal{N}(28.17,\,0.05)^{a}$ & 125.4 & 1.7 \\
\textbf{2-planet} & \textbf{Exterior} & \textbf{2:1} & \textbf{Wide $(-)$} & $\boldsymbol{\mathcal{N}(29.81,\,0.05)}$ & \textbf{123.7} & \textbf{0.0} \\
2-planet & Exterior & 3:2 & Shortward $(+)$ & $\mathcal{N}(21.42,\,0.05)$ & 134.0 & 10.3 \\
2-planet & Exterior & 3:2 & Wide $(-)$ & $\mathcal{N}(22.04,\,0.05)$ & 134.7 & 11.0 \\
2-planet & Exterior & 4:3 & Shortward $(+)$ & $\mathcal{N}(19.13,\,0.05)$ & 132.9 & 9.2 \\
2-planet & Exterior & 4:3 & Wide $(-)$ & $\mathcal{N}(19.50,\,0.05)$ & 130.6 & 6.9 \\
2-planet & Exterior & 5:4 & Shortward $(+)$ & $\mathcal{N}(17.98,\,0.05)$ & 135.4 & 11.7 \\
2-planet & Exterior & 5:4 & Wide $(-)$ & $\mathcal{N}(18.23,\,0.05)$ & 136.1 & 12.4 \\

\midrule
2-planet & Interior & 2:1 & Short-period $(+)$ & $\mathcal{N}(7.14,\,0.05)$ & 128.8 & 5.1 \\
2-planet & Interior & 2:1 & Long-period $(-)$ & $\mathcal{N}(7.35,\,0.05)$ & 134.6 & 10.9 \\
2-planet & Interior & 3:2 & Short-period $(+)$ & $\mathcal{N}(9.57,\,0.05)$ & 135.8 & 12.1 \\
2-planet & Interior & 3:2 & Long-period $(-)$ & $\mathcal{N}(9.75,\,0.05)$ & 136.5 & 12.8 \\
2-planet & Interior & 4:3 & Short-period $(+)$ & $\mathcal{N}(10.79,\,0.05)$ & 136.0 & 12.3 \\
2-planet & Interior & 4:3 & Long-period $(-)$ & $\mathcal{N}(10.94,\,0.05)$ & 136.0 & 12.3 \\
2-planet & Interior & 5:4 & Short-period $(+)$ & $\mathcal{N}(11.52,\,0.05)$ & 136.4 & 12.7 \\
2-planet & Interior & 5:4 & Long-period $(-)$ & $\mathcal{N}(11.65,\,0.05)$ & 136.1 & 12.4 \\
\bottomrule
\end{tabular*}

\vspace{0.5em}
\begin{minipage}{0.98\textwidth}
\footnotesize
$^{a}$ $\mathcal{N}(\mu,\sigma)$ represents a normal distribution with mean $\mu$ and standard deviation $\sigma$.
\end{minipage}
\end{table*}

\begin{table*}[htbp]
\centering
\caption{Priors for the TOI-3850 system used in the joint RV+TTV fit.}
\label{tab:priors}
\vspace{0.5em}
\renewcommand{\arraystretch}{1.15}
\begin{tabular*}{0.95\textwidth}{@{\extracolsep{\fill}}llll@{}}
\toprule
Parameter & TOI-3850 b & TOI-3850 c & Description \\
\midrule

\multicolumn{4}{c}{\textit{Planetary Parameters}} \\
\midrule
$P$ (days) 
& $\mathcal{U}(14,\,15)$ 
& $\mathcal{U}(26,\, 32)^{a}$ 
& Orbital period \\

$T_0$ (TBJD)$^{b}$ 
& $\mathcal{U}(1710,\, 1720)$ 
& $\mathcal{U}(4000,\,4035)$ 
& Transit epoch \\

$K$ ($\mathrm{m~s^{-1}}$) 
& $\mathcal{U}(0,\,50)$ 
& $\mathcal{U}(0,\,50)$ 
& RV semi-amplitude \\

$\sqrt{e}\cos\omega$ 
& $\mathcal{N}(0,0.1)$
& $\mathcal{N}(0,0.1)$ 
& Eccentricity parametrization \\

$\sqrt{e}\sin\omega$ 
& $\mathcal{N}(0,0.1)$ 
& $\mathcal{N}(0,0.1)$ 
& Eccentricity parametrization \\

$i$ (deg) 
& $\mathcal{N}(88.44,\,0.01)$ 
& $\mathcal{U}(70,\,90)$ 
& Orbital inclination \\

$b$ 
& -- 
& $>1$ 
& Impact parameter \\

$\Omega$ (deg) 
& 0 (fixed) 
& $\mathcal{N}(0,\,15)$ 
& Longitude of the ascending node \\

\midrule
\multicolumn{4}{c}{\textit{RV Parameters}} \\
\midrule
$\mu_{\rm MAROON\mbox{-}X}$ ($\mathrm{m~s^{-1}}$) 
& \multicolumn{2}{c}{$\mathcal{U}(-50,\,50)$} 
& MAROON-X RV offset \\

$\sigma_{\rm MAROON\mbox{-}X}$ ($\mathrm{m~s^{-1}}$) 
& \multicolumn{2}{c}{$\mathcal{U}(0,\,50)$} 
& MAROON-X RV jitter \\

\bottomrule
\end{tabular*}

\vspace{0.5em}
\begin{minipage}{0.95\textwidth}
\footnotesize
$^{a}$ $\mathcal{U}(a,b)$ denotes a uniform distribution bounded between $a$ and $b$. \\
$^{b}$ TBJD is BJD $-2457000$. \\
\end{minipage}

\end{table*}

\bibliography{refs}{}
\bibliographystyle{aasjournalv7}

\end{document}